%% file: main.tex
\documentclass[fleqn,usenatbib]{mnras}

\usepackage{newtxtext,newtxmath}

\usepackage[T1]{fontenc}

\DeclareRobustCommand{\VAN}[3]{#2}
\let\VANthebibliography\thebibliography
\def\thebibliography{\DeclareRobustCommand{\VAN}[3]{##3}\VANthebibliography}

\usepackage{graphicx}	
\usepackage{amsmath}	
\usepackage{multirow}
\usepackage{tabularray}

\usepackage{etex} 
\usepackage{enumitem}
\usepackage{enumerate}

\title[Dark Matter Searches]{Probing dark matter with adaptive-optics based flux ratio anomalies: photometric and astrometric precision}

\author[I. A. Zelko]{
Ioana A. Zelko,$^{1,2}$\thanks{Corresponding author, e-mail: ioana.zelko@gmail.com}
Anna M.~Nierenberg,$^{2}$
Tommaso Treu$^{3}$
\\
$^{1}$Canadian Institute for Theoretical Astrophysics, University of Toronto, 60 St George Street, Toronto, M5S 3H8, Ontario, Canada\\
$^{2}$Department of Physics and Astronomy, University of California-Los Angeles,
475 Portola Plaza, Los Angeles, CA 90095, USA\\
$^{3}$University of California Merced, Department of Physics 5200 North Lake Rd. Merced, CA 9534\\
}

\pubyear{2023}

\begin{document}
\label{firstpage}
\pagerange{\pageref{firstpage}--\pageref{lastpage}}
\maketitle

\begin{abstract}

%

Strong gravitational lensing is a powerful probe of the distribution of matter on sub-kpc scales. It can be used to test the existence of completely dark subhalos surrounding galaxies, as predicted by the standard cold dark matter model, or to test alternative dark matter models. The constraining power of the method depends strongly on photometric and astrometric precision and accuracy. We simulate and quantify the capabilities of upcoming adaptive optics systems and advanced instruments on ground-based telescopes, focusing as an illustration on the Keck Telescope (OSIRIS + KAPA, LIGER + KAPA) and the Thirty Meter Telescope (TMT; IRIS + NFIRAOS). We show that these new systems will achieve dramatic improvements over current ones in both photometric and astrometric precision. Narrow line flux ratio errors below $2\%$, and submilliarcsecond astrometric precision will be attainable for typical quadruply imaged quasars. With TMT, the exposure times required will be of order a few minutes per system, enabling the follow-up of 100-1000 systems expected to be discovered by the Rubin, Euclid, and Roman Telescopes.
\end{abstract}



\input{instrument_specs.tex}

\section{Introduction}

%

Multiple lines of evidence suggest that the majority of the mass in the Universe is non-baryonic (i.e., not quarks) and does not interact with light. Understanding the fundamental nature of this so-called dark matter has proven elusive so far. 

In the standard cosmological model, dark matter is assumed to be "cold", i.e., made of non-relativistic particles that do not have significant interactions with each other or with baryonic matter, apart from gravity. One example of cold dark matter is Weakly Interactive Massive Particles (WIMPs). This CDM model is extremely successful at reproducing observations of the universe at scales larger than an Mpc, including e.g., the cosmic microwave anisotropy, baryonic acoustic oscillations, and light element abundances.

However, many alternative dark matter models have been proposed both on theoretical grounds and as a means to solve some of the tensions that have been identified at galactic and subgalactic scales \citep[see][for a review and references therein]{BB17}. These alternative models can be classified in terms of their particle physics properties, such as formation mechanisms and interactions, or in terms of their astrophysical signatures. 
In astrophysical contexts, 'warm' refers to models with a non-negligible free-streaming length, while 'self-interacting' pertains to those with significant dark matter self-interaction cross-sections.

Increasingly sensitive ground-based direct detection experiments have been able to set limits to the cross section of interaction with regular matter. In parallel, a variety of astronomical observations have constrained some of the generic properties of dark matter, based on electromagnetic observations. For example, the mass function of Milky Way Satellites and the non-detection of them in the Gamma and X-ray sets limits on the free streaming lengths and self-interaction cross section \cite[e.g.,][and references therein]{Strigari18}. 

In recent years, strong gravitational lensing has emerged as a powerful complement to these traditional techniques to study the nature of dark matter, because it relies exclusively on gravity, the only known interaction of dark matter. Dark matter models make very specific predictions on how mass should be distributed in the universe.  Generically, dark matter models, together with appropriate models of non-linear growth and baryonic effects, can predict the clumping of dark matter as a function of scale, known as the halo and subhalo mass function,  as well as the internal structure of halos and subhalos. CDM predicts a power law halo mass function rising as M$^{-1.9}$ towards small masses \citep{Springel08}, with dark matter density profiles described by the \citet{NFW97} profile, subject to modifications due to baryonic effects.

By carefully measuring the distortion of the multiple images of distant sources, gravitational lensing probes the foreground distribution on subgalactic scales, thus allowing scientists to perform a direct test of dark matter model predictions. For example, in CDM dark matter, halos and subhalos below $10^7$M$_\odot$ should be extremely abundant yet dark because their potential wells are not deep enough to retain gas and stars. Strong lensing has the potential to test this fundamental prediction of the model, since it is sensitive to substructure independent of its electromagnetic emission.

In practice, a powerful way to constrain dark matter models using strong lensing is to measure the flux ratios of unresolved, multiply-imaged sources, often referred to as the "flux ratio anomaly" technique. The flux ratios between multiple images depend on a combination of second derivatives of the projected gravitational potential at the location of the images, making them very sensitive to the local "clumpiness" of matter. A cold dark matter universe will have many more halos and subhalos near a multiple image than a universe with warm dark matter, and thus results in more "anomalous" flux ratios with respect to those predicted by a smooth mass distribution \citep{M+S98,Cyr-Racine19}. 

One complication is that in typical systems, composed of a galaxy lensing a background quasar, the stars in the foreground galaxy can also affect the flux ratios, a phenomenon called optically thick microlensing\footnote{Because the Einstein radius of a star is of order microarcseconds at the distances under consideration}. Thus, in order to filter this signal out one needs to study sources that are in projection significantly larger than microarcseconds to smooth over the stellar microlensing. For this reason, the first papers on this topic focused on radio-loud lensed quasars \citep{D+K02,D+K06,M+M01}. More recently, with the advent of adaptive optics-assisted integral field spectrographs and spectrographs in space, it has become possible to use the narrow line region of quasars as a tracer \citep{M+M03,Nie++14,Nie++17,Nierenberg2020}. The mid-infrared emission of the hot dust surrounding the AGN offers another exciting window to measure flux ratios. Although only detectable in a limited number of instances with 8-10m class telescopes \citep[][]{Chiba2005}, this technique is well within the capabilities of the James Webb Space Telescope (JWST) \citep{Nierenberg2023} and will be accessible to Extremely Large Telescopes in the future. 


In addition to affecting the flux ratios of multiple images, the distribution of matter on small scales also influences their positions, creating what are known as "astrometric anomalies" at the milliarcsecond level \citep{Chen2007}. There are other approaches that exploit both the 'flux ratio anomalies' and the 'astrometric anomalies', by themselves or in combination with the distortion observed in multiply-imaged, extended sources \citep{Vegetti09,Birrer2021}. These methodologies provide powerful tools for understanding dark matter and galactic structure \citep{Birrer23}.

The sensitivity of strong lensing as a probe of dark matter in each observed system relies on angular resolution and sensitivity. These factors are crucial as they determine the relative photometric precision and accuracy, as discussed by \citet{gilman2019b}, along with the relative astrometric precision and accuracy, which is explored by \citet{Chen2021}. In addition, the statistical power of the method increases with the number of systems that one can study, because by probing a larger volume one averages over the stochastic nature of the halo and subhalo distribution.

For these reasons, adaptive optics on large ground-based optical infrared telescopes can be transformative, enabling the observations of a large number of lensing systems with sufficient photometric and astrometric precision. Beyond the obvious sensitivity and resolution requirements, the key issues with laser guide star AO will be sky coverage (to maximize targets) and the ability to reconstruct the point spread function (PSF) well enough to achieve the desired precision and accuracy. 

In this paper, we investigate the performance of current and future laser guide star adaptive optics systems, by carrying out detailed simulations of flux and astrometric anomalies as measured from integral field spectroscopy of lensed narrow emission lines. Even though there is no hard threshold and every improvement in precision helps, it is clear based on \citet{gilman2019b} that percent level photometry is needed and that milliarcsecond (mas) level astrometry is needed to make significant progress over what we know so far, for large numbers of systems. Therefore we set as our target requirements 2\% total error in flux ratio photometry, and mas total error in astrometry, and show that they are attainable. We also discuss the total amount of exposure time required per system to give a rough estimate of the total amount of time required to investigate certain dark matter models, even though in practice this will depend on the distribution of fluxes of the actual lenses to be observed.


In Section \ref{sec:observatories} we detail the instruments considered, their PSFs. In Section~\ref{sec:mocklenses} we describe how we model the quadruply imaged systems, and the simulation cases considered. Section \ref{sec:inference} goes over our inference methodology and pipeline. We present our results in Section \ref{sec:results}.
%
%
%
%
%
%
%
%

\begin{figure*}
	\begin{center}\hspace{-5mm}
 KAPA \hspace{7.7cm} NFIRAOS\\
		\begin{tabular}{cccc}
			\hspace{-5.20mm}
			\includegraphics[width=0.5\textwidth]{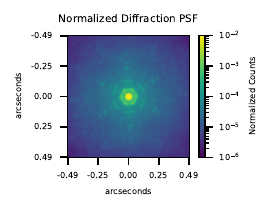} &
			\hspace{-4.70mm}
			\includegraphics[width=0.5\textwidth]{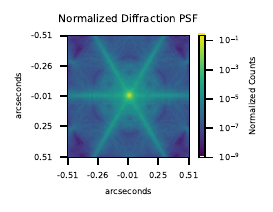}
		\end{tabular}
		\caption{Diffraction limited point-spread functions, displayed on a logarithmic scale for the KAPA adaptive optics system (left panel), and for the NFIRAOS adaptive optics system (right panel). The hexagonal pattern created by the mirror/telescope structure is apparent. \label{fig:psf_diffraction}}
	\end{center}
\end{figure*}

\begin{figure*}
	\begin{center} \hspace{-6mm}
 Current AO \hspace{5.3cm} KAPA \hspace{5.3cm} NFIRAOS
		\begin{tabular}{cccc}
			\hspace{-11.20mm}
			
			\includegraphics[width=0.37\textwidth]{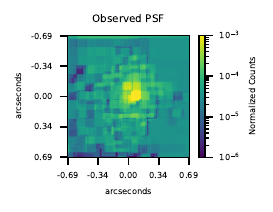} &
			\hspace{-7.70mm}\includegraphics[width=0.37\textwidth]{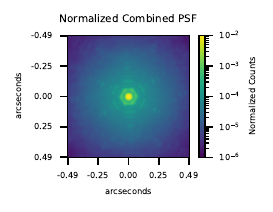} &
			\hspace{-6.70mm}
			\includegraphics[width=0.37\textwidth]{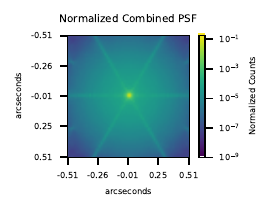}
		\end{tabular}
		\caption{Effective point spread functions in typical observing conditions. Left: current adaptive optics system as measured (see Section OSIRIS PSF). Center: KAPA adaptive optics system, simulated. Right: NFIRAOS adaptive optics system, simulated. The two simulated PSFs assume seeing of  0\farcs3, and a Strehl ratio of 0.7.}{ \label{fig:psf}}
	\end{center}
\end{figure*}

\begin{figure*}
	\begin{center}
  KAPA \hspace{+7.5cm} NFIRAOS

		\begin{tabular}{cccc}
			\hspace{-5.20mm}
			\hspace{-5.20mm}
			\includegraphics[width=0.5\textwidth]{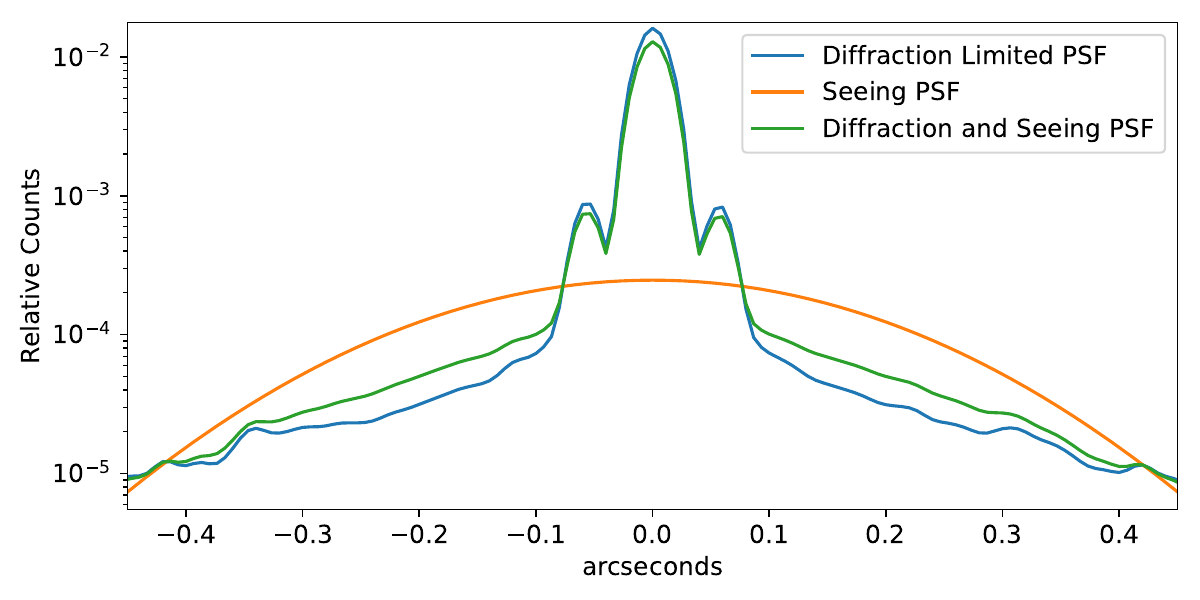} &
			\hspace{-4.70mm}
			\includegraphics[width=0.5\textwidth]{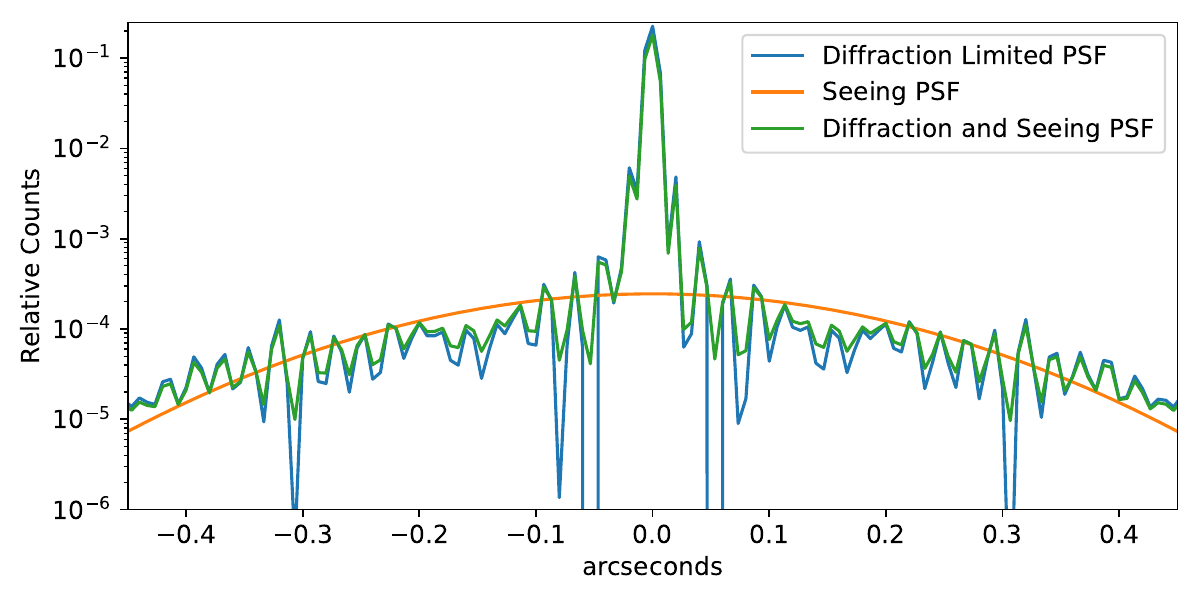}
			
		\end{tabular}
		\caption{Cross-section of the point spread function for the KAPA adaptive optics system (left panels), and for the NFIRAOS adaptive optics system (right panels). They are presented in linear scale in the upper panels, and in Log scale in the lower panels. The blue curves represent the diffraction-limited PSFs, the orange ones the uncorrected seeing, and the green ones the combined PSFs assuming a Strehl of 0.8 and seeing FWHM of 0\farcs4. }{ \label{fig:psf_section}}
	\end{center}
\end{figure*}

\section{Observatory characteristics}\label{sec:observatories}

One of our goals is to compare the expected performance of future instrument and adaptive optics systems with current facilities.  For this purpose, in this section, we start with an overview of state of the art and upcoming cutting-edge adaptive-optics system, at the Keck Telescope and the Thirty Meter Telescope(TMT), that we take as illustrative of 8-10 and 30m class capabilities. We also describe our models for the point spread functions and expected noise levels. 


\subsection{Instruments and Adaptive Optics Systems}

Here we provide details of the systems we are comparing/simulating. Table~\ref{table:instrument_specs} compiles the key features of the systems.

{\bf Keck OSIRIS/Current AO} Current dark matter flux ratio measurements are performed with the LGSAO system \citep{STRAP} and the OSIRIS spectrograph \citep{OSIRIS}. LGSAO/OSIRIS observations are the baseline with which we will compare the future facilities.

{\bf Keck OSIRIS/KAPA} Keck's current AO system is currently being upgraded to the Keck All-sky Precision Adaptive-optics \citep[KAPA,][]{wizinowich}, which will have an enhanced Strehl and sky coverage.

{\bf Keck LIGER/KAPA} The existing OSIRIS spectrograph is planned to be upgraded to the LIGER spectrograph \citep{Wright2019}. LIGER will have 4 times the pixels of OSIRIS, a reduced pixel size, and a lower detector dark current and readout noise, improving resolution and sensitivity.

{\bf TMT IRIS/NFIRAOS}  At first light, TMT will feature NFIRAOS, an advanced AO system designed for its larger aperture \citep{NFIRAOS}.
TMT will use IRIS \citep{Wright16}, which is similar to LIGER in design, technology, and performance.

\subsection{Point Spread Functions}
The astrometric quality of images is fundamentally constrained by two main factors: the diffraction by the telescope and its optics, represented by a diffraction point spread function (PSF), and the blurring effect of turbulence in the Earth's atmosphere, represented by the seeing PSF \footnote{Imperfect optics within the instrument can further degrade the PSF, but we will neglect this component here to simplify the discussion.}. 
The total PSF, which describes the response of the imaging system as a whole, is a combination of these two components. 




The efficacy of an imaging system can be quantitatively assessed using the Strehl ratio; a metric that compares the peak intensity of the seeing-aberrated point-source image to that of the ideal diffraction-limited image. Within the context of the total PSF, the Strehl ratio serves as an empirical measure of the optical system's performance. It essentially quantifies how closely the real-world optical system approaches the theoretical performance of a perfect system under identical observing conditions. This is particularly pertinent in adaptive optics, where real-time adjustments are employed to mitigate atmospheric distortions, aiming to achieve near diffraction-limited observations.

%
%
%
%
%

We obtained the diffraction limited PSFs for KAPA (Keck) and NIFARAOS (TMT) from private communications with their respective team members \footnote{Prof. Shelley Wright and Dr. Nils Rundquist}. 

Our current Keck PSF model is obtained from an observation taken under typical seeing conditions with the OSIRIS spectrograph in Hbb with the 100 mas pixel scale. Even though this pixel scale undersamples the PSF, this configuration is the most typical for measuring flux ratios of gravitationally lensed quasars, owing to field of view and signal to noise ratio constraints. The measured PSF shown here is an inferred sub-sampled PSF created using {\tt lenstronomy}\citep{birrer2018a, birrer2021c}.

Figure~\ref{fig:psf_diffraction} depicts the diffraction-limited PSFs for the KAPA (Keck) and NFIRAOS (TMT) systems. Fig.~ \ref{fig:psf} compares the total (diffraction, instrumental, and seeing) PSFs of Keck's current AO system, KAPA, and NFIRAOS, under typical conditions. Fig.~\ref{fig:psf_section} presents cross sections of the diffraction, seeing, and total simulated PSFs for KAPA and NFIRAOS, illustrating the contributions of the subcomponents.

%

\subsection{Noise}
The observed noise is a combination of instrument noise and sky/background noise. The sky/background noise is expected to be the same for all three facilities, while the instrument noise will vary based on the characteristics of the telescope and instrument.

The OSIRIS/LGSAO noise levels are taken directly from real observations in typical conditions. For Liger and IRIS, we simulate the noise as a combination of instrument noise and sky /background. The instrument noise is determined by the detector technical specifications (refer to Table~ \ref{table:instrument_specs}), and the sky background based on the OSIRIS/LGSAO real observations after subtracting the estimated contribution of instrument level effects. For our noise calculations we assume 600 second exposure times, which is the exposure time used for OSIRIS/LGSAO observations of this kind of systems \citep{Nierenberg14}. Details for the noise calculations are described in Appendix ~\ref{sec:app_noise}.





%
%

\begin{table*}\label{table:quad_prop}
	\caption{Positions of the main deflector's light profile centroid and lensed QSO images; quasar image magnitudes in the AB system; model parameters for the lens mass distribution.}
	\setlength{\tabcolsep}{0.1pt}
	\begin{tabular}{c|c|c|c|c|c|c|c|c|c|c|c|}
		\hline  $\begin{array}{l}\text { Type of } \\
			\text { lens system }\end{array}$ &  $\begin{array}{l}\text { Name of } \\
			\text { lens system }\end{array}$ & \multicolumn{2}{|c|}{ Main deflector } & \multicolumn{2}{|c|}{ Image A } & \multicolumn{2}{|c|}{ Image B } & \multicolumn{2}{|c|}{ Image C } & \multicolumn{2}{|c|}{ Image D } \\
		& & $\begin{array}{c}\Delta \mathrm{RA} \\
			(\operatorname{arcsec})\end{array}$ & $\begin{array}{c}\Delta \text { Dec. } \\
			(\operatorname{arcsec})\end{array}$ & $\begin{array}{c}\Delta \mathrm{RA} \\
			(\operatorname{arcsec})\end{array}$ & $\begin{array}{c}\Delta \text { Dec. } \\
			(\operatorname{arcsec})\end{array}$ & $\begin{array}{c}\Delta \mathrm{RA} \\
			(\operatorname{arcsec})\end{array}$ & $\begin{array}{c}\Delta \text { Dec. } \\
			(\operatorname{arcsec})\end{array}$ & $\begin{array}{c}\Delta \mathrm{RA} \\
			(\operatorname{arcsec})\end{array}$ & $\begin{array}{c}\Delta \text { Dec. } \\
			(\operatorname{arcsec})\end{array}$ & $\begin{array}{c}\Delta \mathrm{RA} \\
			(\operatorname{arcsec})\end{array}$ & $\begin{array}{c}\Delta \text { Dec. } \\
			(\operatorname{arcsec})\end{array}$ \\
		\hline 
		\text{Cross}&\text { ATLAS J2344-3056 } & 0.055 & -0.115 & -0.431 & 0.073 & 0.150 & 0.425 & 0.442 & -0.245 & -0.191 & -0.583\\
		\text{Cusp}&\text { J2205-3727 } & -0.046 & 0.088 & 0.859 & 0.189 & -0.342 & -0.536 & -0.780 & 0.066 & -0.491 & 0.624 \\ 
		\text{Fold}&\text { J1131-4419 } & 
		-0.012 & 0.050 & 0.072 & -0.742 & -0.881 & 0.349 & 0.411 & 0.793 & 0.754 & 0.446 \\ 
		\hline
	\end{tabular}
	\setlength{\tabcolsep}{1pt}
	
	\begin{tabular}{llccccc}
		\hline Type of lens & Name of lens & Filter & A (mag) & B (mag) & C (mag) & D (mag)  \\
		\hline 
		Cross  & PS J0630-1201 & $F 160 W$  & 21.253 & 21.097 & 20.421 &  21.303 \\
		\text{Cusp}&\text { J2205-3727 }  & $F 160 W$& 22.183 & 21.643 & 21.222 & 22.494\\ 
		\text{Fold}&\text { J1131-4419 }  & $F 160 W$ & 20.317 &19.993 & 19.420 & 19.235\\ 
		\hline
	\end{tabular}
	
	\begin{tabular}{llccccc}
		\hline $\begin{array}{l}\text { Name of } \\
			\text { lens system }\end{array}$ & $\begin{array}{l}\text { Name of } \\
			\text { lens system }\end{array}$ & $n$ & $\begin{array}{c}\theta_{\mathrm{e}} \\
			(\operatorname{arcsec})\end{array}$ & $q_{\mathrm{L}}$ & $\begin{array}{c}\phi_{\mathrm{L}}(\mathrm{N} \text { of E) } \\
			\left({ }^{\circ}\right)\end{array}$ &  $\begin{array}{c}I_{\mathrm{e}}(F 160 W) \\
			\left(\mathrm{mag}  \mathrm{\ arcsec}^{-2}\right)\end{array}$ \\
		\hline
		\text{Cross}&\text { ATLAS J2344-3056 } & 3.60& 1.45& 0.84& -22.9& 21.38\\
		\text{Cusp}&\text { J2205-3727 } & 6.60 &0.78& 0.74 &-76.7& 20.84 \\ 
		\text{Fold}&\text { J1131-4419 } & 4.0 &0.29& 0.61& 83.4& 18.31 \\ 
		\hline
	\end{tabular}
	
\end{table*}

\begin{table*}
	\centering
	\caption{Summary of the simulations carried out}
	\begin{tabular}{|l|llll}
		\hline
		\textbf{Simulation Parameter} & \textbf{Parameter Values}	\\
		\hline
		Observational Setup                                            & Keck OSIRIS+Current AO              & Keck OSIRIS+KAPA      & Keck LIGER+KAPA                            & \multicolumn{1}{|l|}{TMT IRIS+NFIRAOS} \\ \hline
		Quad lens systems                                              & Cross: ATLASJ2344-3056         & Cusp: J2205-3727 & \multicolumn{1}{l|}{Fold: J1131-4419} &                                   \\ \cline{1-4}
		PSF                                                         & Assumed Known & Assumed Unknown                  &                                       &                                   \\ \cline{1-3}\\
		Main deflector light model                                                         & Included & Not included                 &                                       &                                   \\ \cline{1-3}\\
		Seeing FWHM  (arcsec) & \multicolumn{1}{l|}{0.3 - 0.8} &                  &                                       &                                   \\ \cline{1-2}
		Strehl                                                         & \multicolumn{1}{l|}{0.3 - 0.9} &                  &                                       &                                   \\ \cline{1-2}\\
		
	\end{tabular}
	
	\label{table:simulation_table}
\end{table*}

\section{Strong gravitational lens simulations}\label{sec:mocklenses}

Here we describe how we generate the mock gravitational lensing systems for our analysis. We begin by describing the lensing configurations we choose. Then we describe how the different light components are modelled. Finally we describe the combination of lenses and instrumental setups that we simulate. 

We consider three cases for gravitational quad systems, corresponding to the three most common configurations: a cross, a cusp, and a fold. We adopt as representatives of these classes the real systems modeled by \cite{schmidt2023}, ATLASJ2344-3056  (cross), J2205-3727 (cusp), J1131-4419 (Fold). 
Table \ref{table:quad_prop} lists the parameters of the quads: image positions, magnitudes, and lens galaxy light profile.

In order to produce our estimate of astrometric and flux ratio precision and accuracy we start from "white light" images, obtained by integrating data based over the range corresponding to the rest-frame 4700 and 5100 Å. This range contains the strong H$\beta$ and [\ion{O}{III}] emission lines generally used for flux ratio anomalies, and that we expect will be the main workhorse to follow-up lenses discovered by LSST, Euclid, and Roman \cite{O+M10}.  We convert \emph{HST} F160W band magnitudes measured by \citet{schmidt2023} to predicted detector counts using the ratio of the observed magnitudes and OSIRIS detector counts for gravitational lenses observed by Nierenberg et al. (2024, in prep). 

The quasar point sources are inserted directly into the images using the appropriate point spread function for the selected AO/instrument combination.




We test the effect of the deflector light on the accuracy of the inference by including the deflector for some of our mock systems. We model the deflector as an elliptical Sersic with equation given by: 

\begin{equation}
I(\theta)=I\left(\theta_{\mathrm{e}}\right) \exp \left\{-C(n)\left[\left(\frac{\left(q_{\mathrm{L}} \theta_1\right)^2+\theta_2^2}{q_{\mathrm{L}} \theta_{\mathrm{e}}^2}\right)^{\frac{1}{2 n}}-1\right]\right\},
\label{eq:sersic}
\end{equation}

where $C(n)$ is a normalization constant so that at the effective radius, $\theta_{\mathrm{e}}$, the profile includes half of the deflector's light. $n$ represents the Sérsic index, $\theta_1$ and $\theta_2$ are the angular coordinates aligned along the semi-major and semi-minor axis through the rotational PA $\phi_{\mathrm{L}}=\arctan \left(\theta_2, q \theta_1\right)$ of the light profile, and $q_{\mathrm{L}}$ represents the corresponding axial ratio.  See Appendix~\ref{sec:lens_conversion} for an explanation of how to convert these parameters to the definitions used within\textit{lenstronomy}. 

Each of the three exemplar lens systems are simulated with every instrumental setup, with and without the light from the deflector, for a range of seeing values. During the inference, as discussed in the next section we consider two cases: a known PSF (from, e.g., telemetry and modeling) and an unknown PSF that needs to be reconstructed from the data. Table~\ref{table:simulation_table} summarizes the key properties of the simulated systems.

\begin{figure*}
	\begin{center}
		\begin{tabular}{cccc}
			\hspace{-12mm}
			\includegraphics[width=0.3\textwidth]{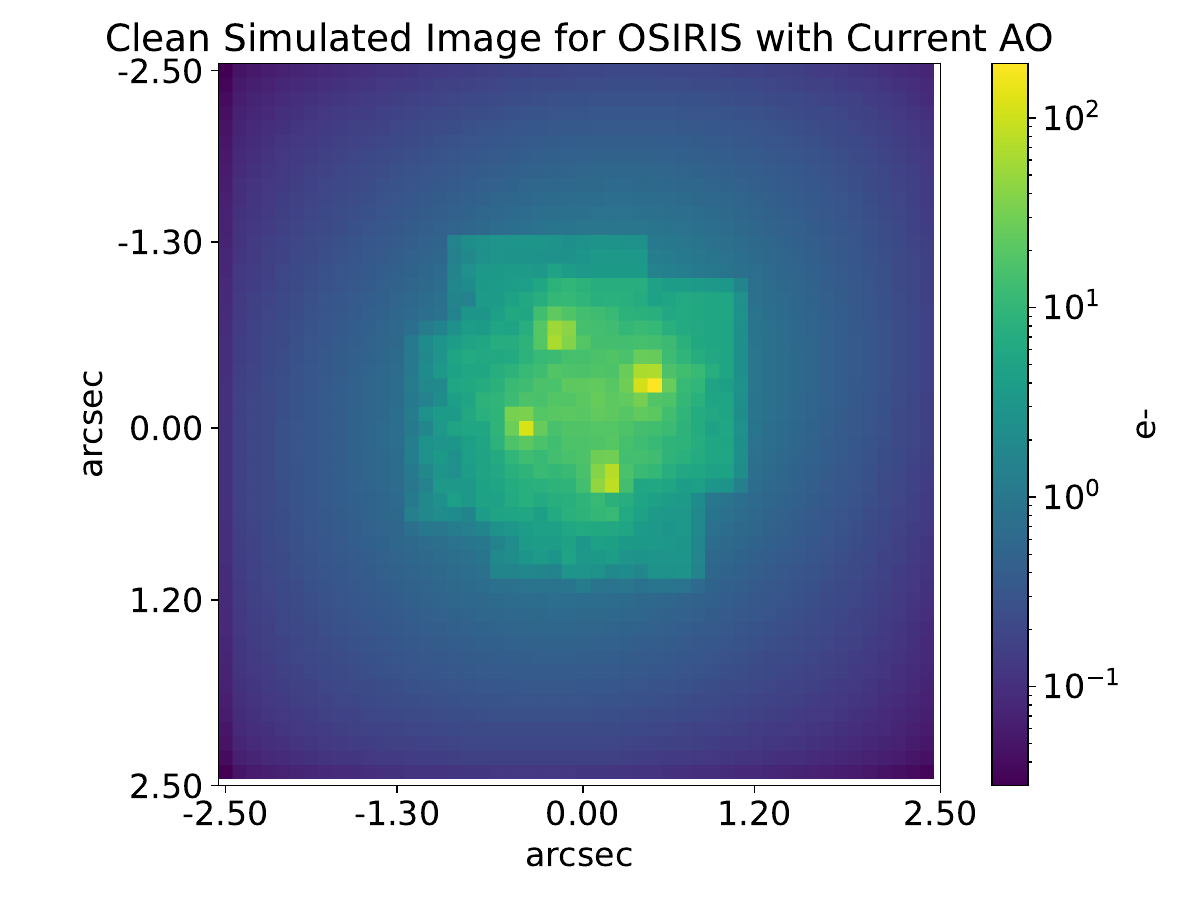} &
			\hspace{-9.2mm}
			\includegraphics[width=0.3\textwidth]{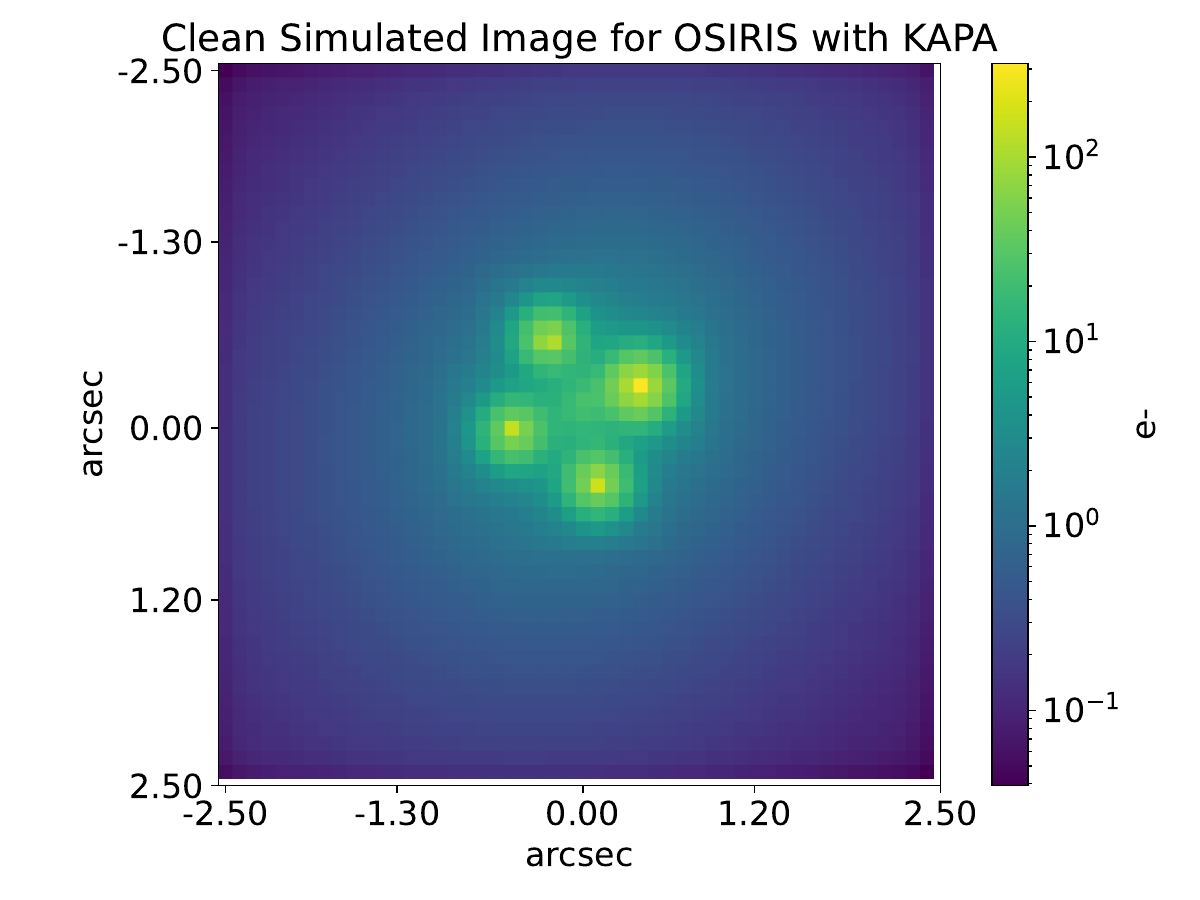} &
			\hspace{-9.2mm}
			\includegraphics[width=0.3\textwidth]{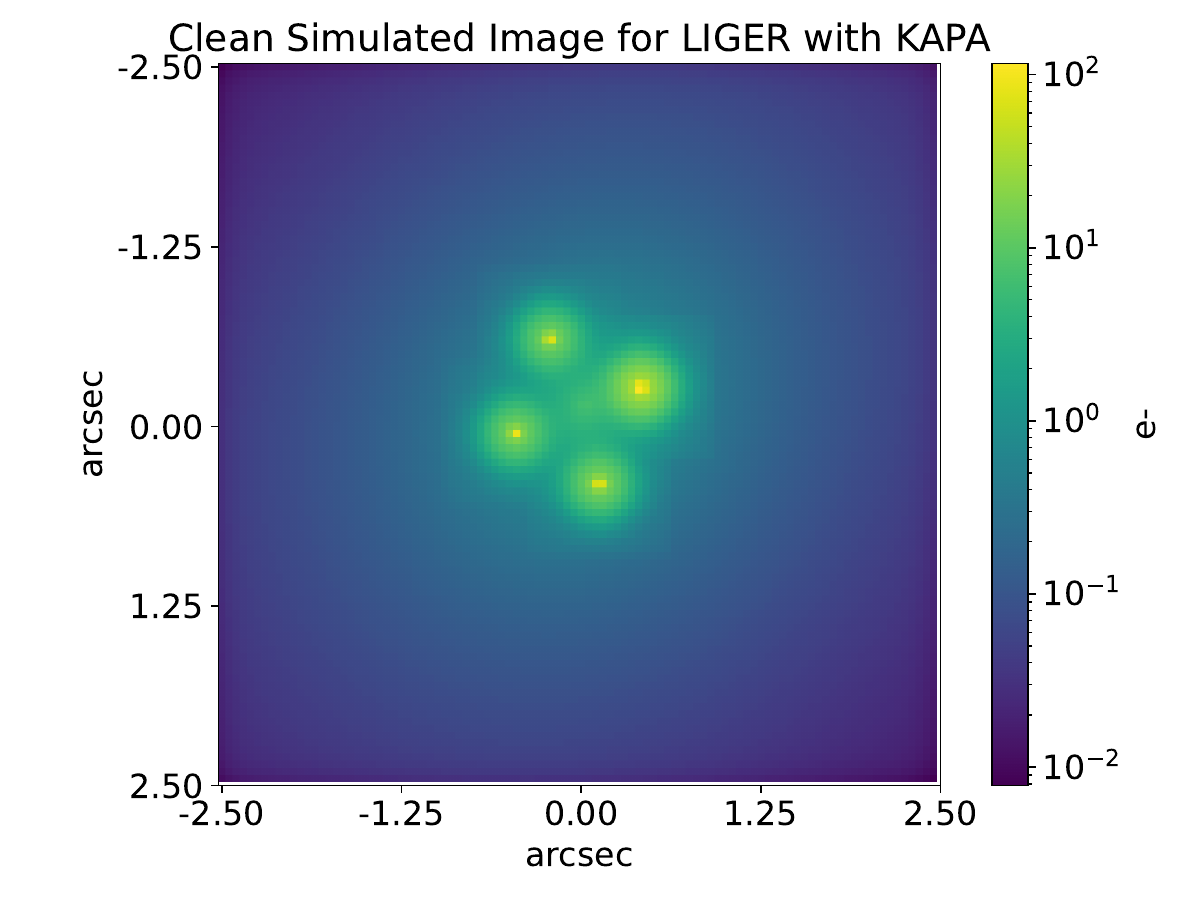} &
			\hspace{-9.2mm}
			\includegraphics[width=0.3\textwidth]{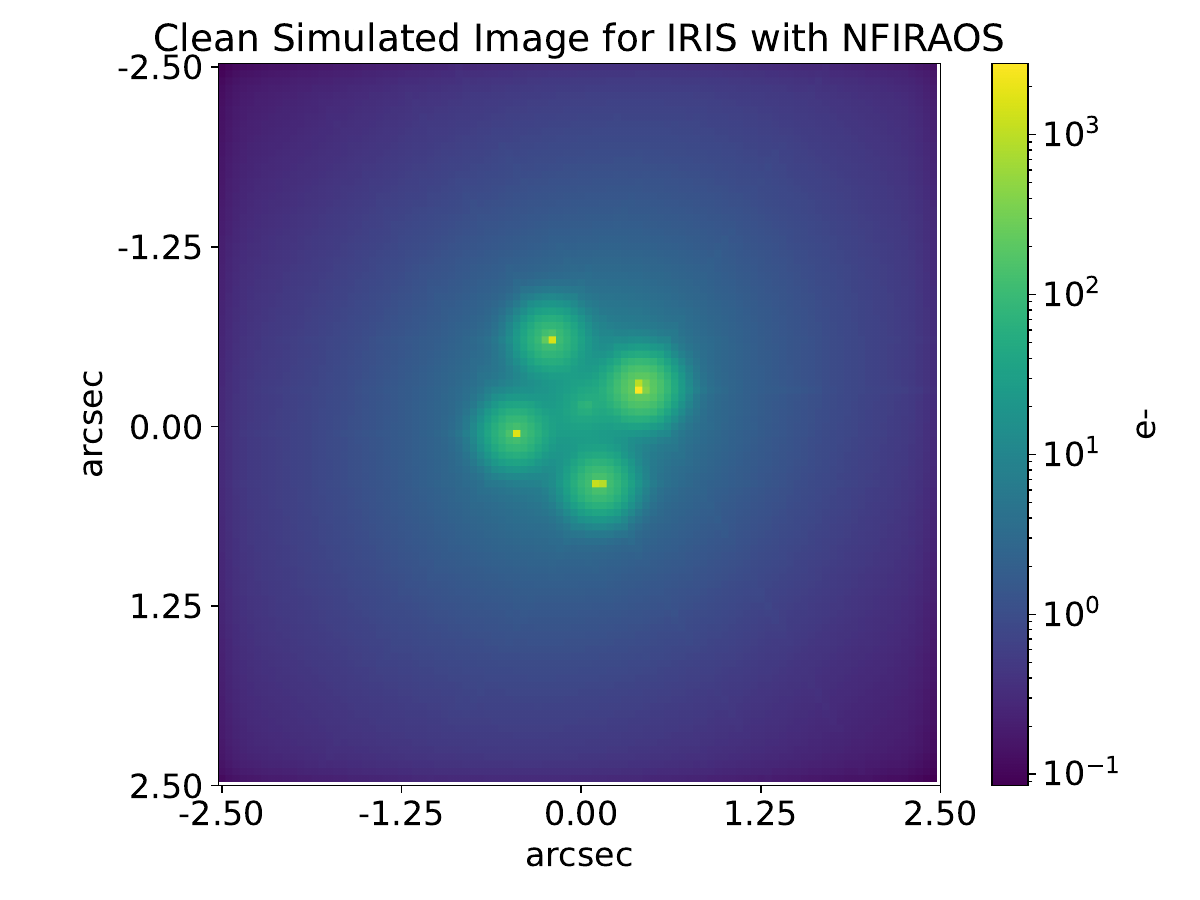} \\
			\hspace{-12mm}
			\includegraphics[width=0.3\textwidth]{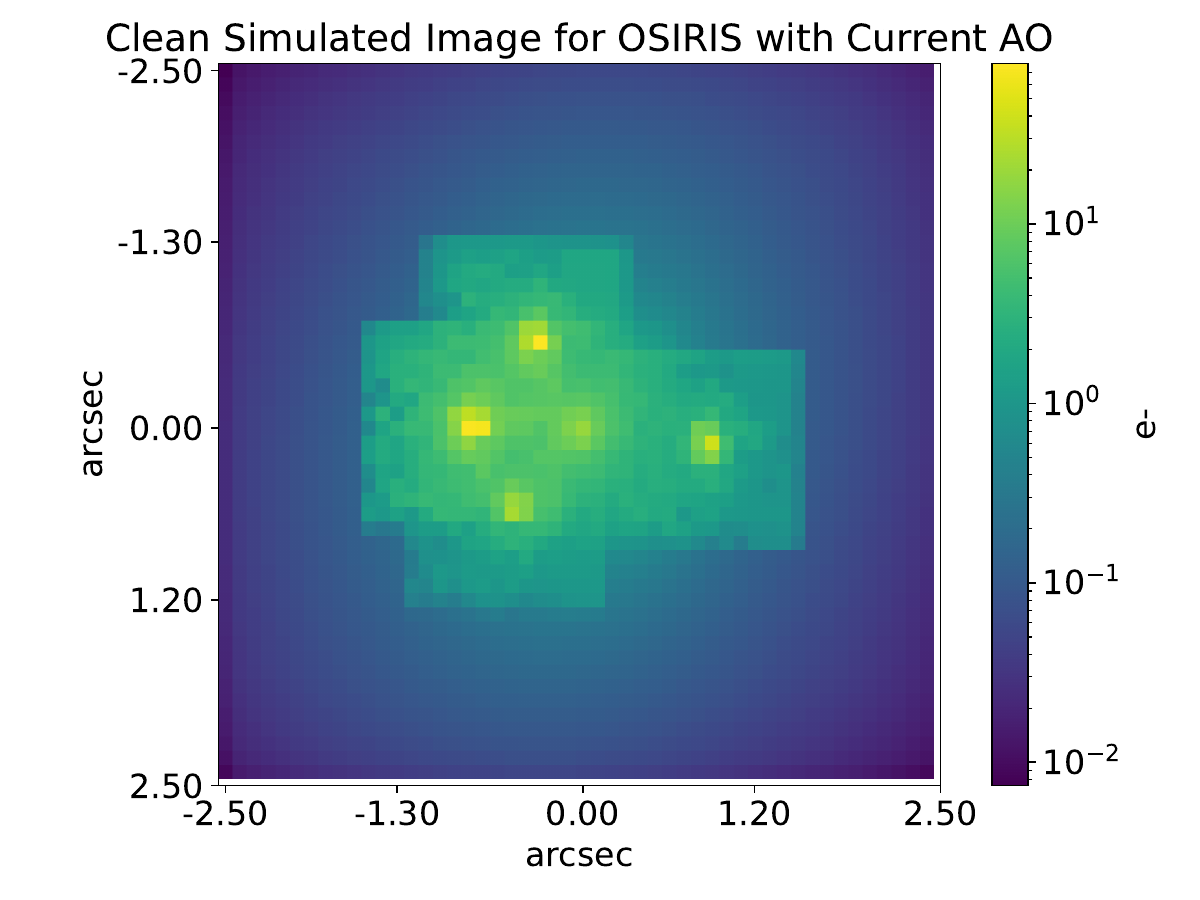} &
			\hspace{-9.2mm}
			\includegraphics[width=0.3\textwidth]{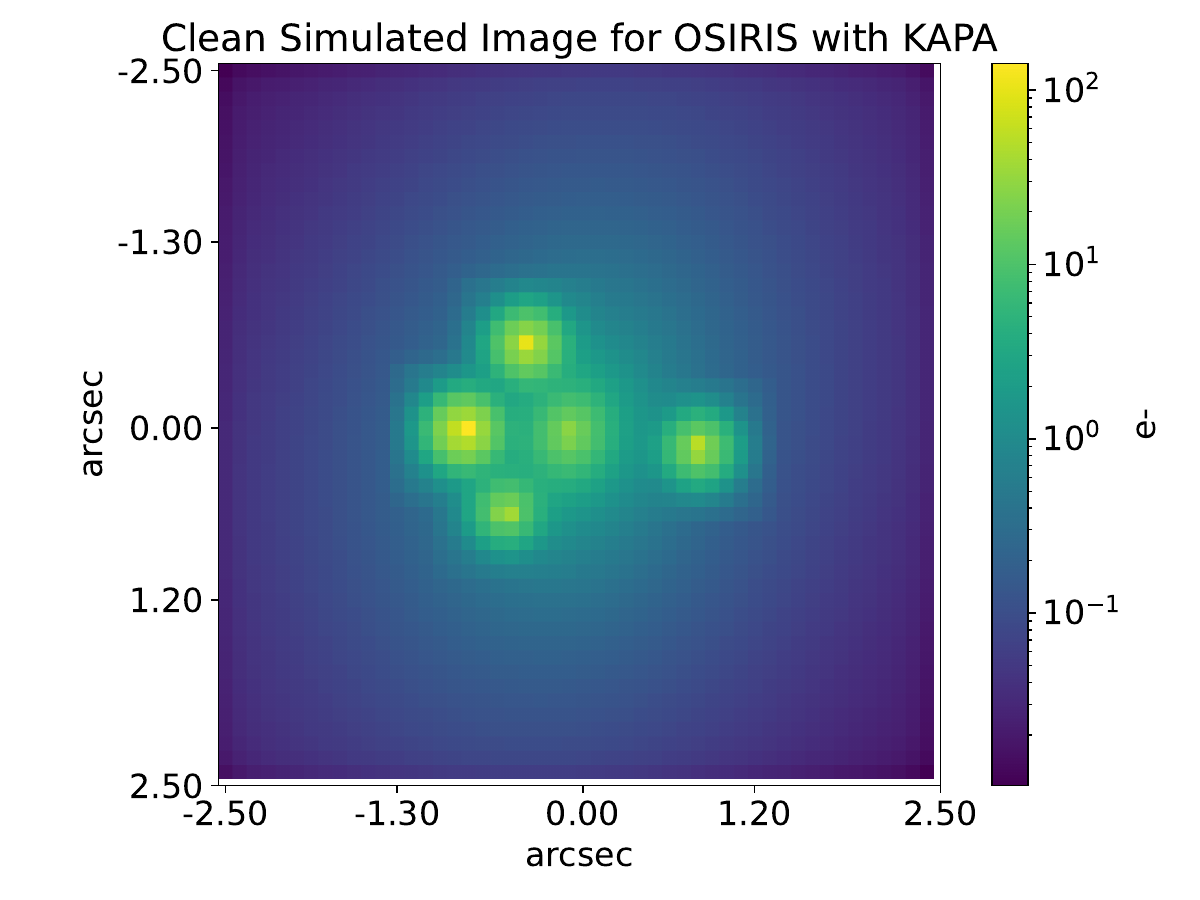} &
			\hspace{-9.2mm}
			\includegraphics[width=0.3\textwidth]{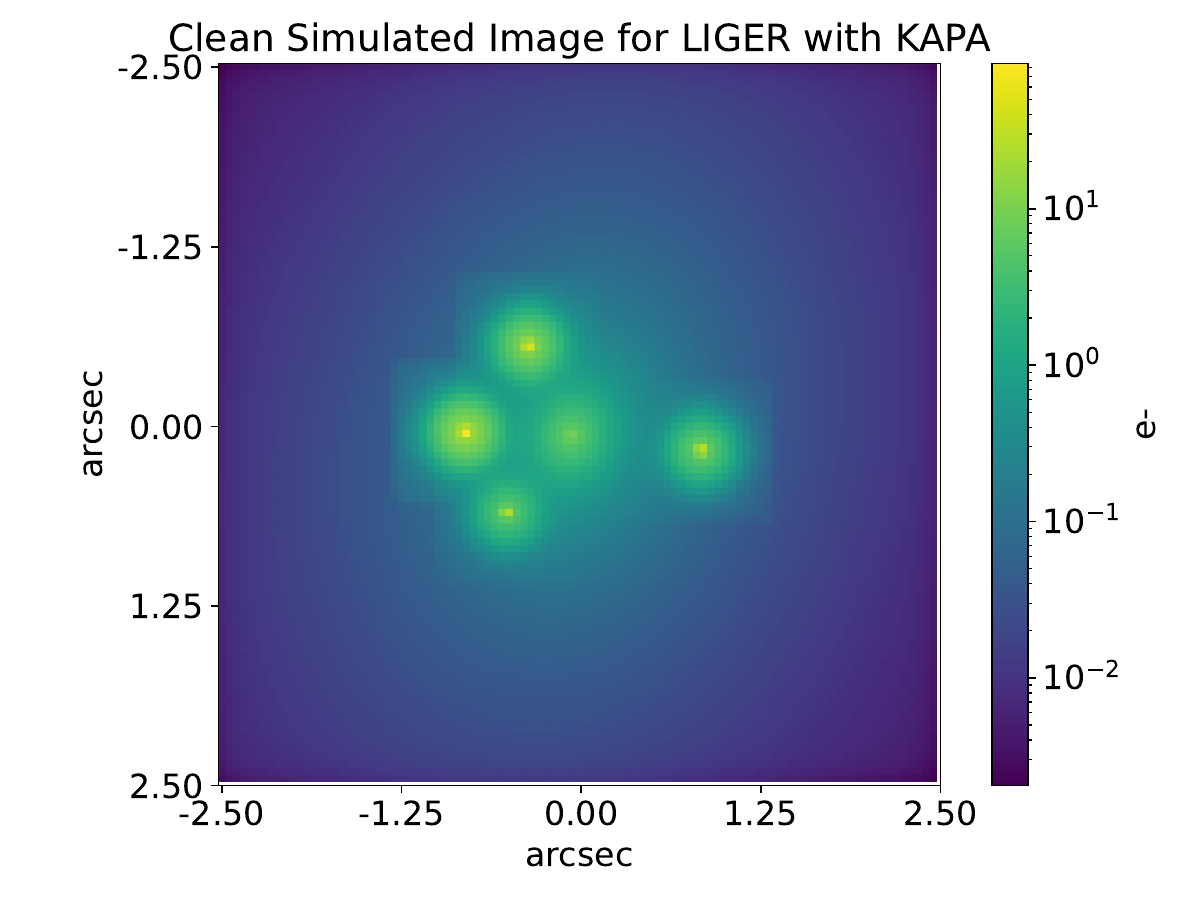} &
			\hspace{-9.2mm}
			\includegraphics[width=0.3\textwidth]{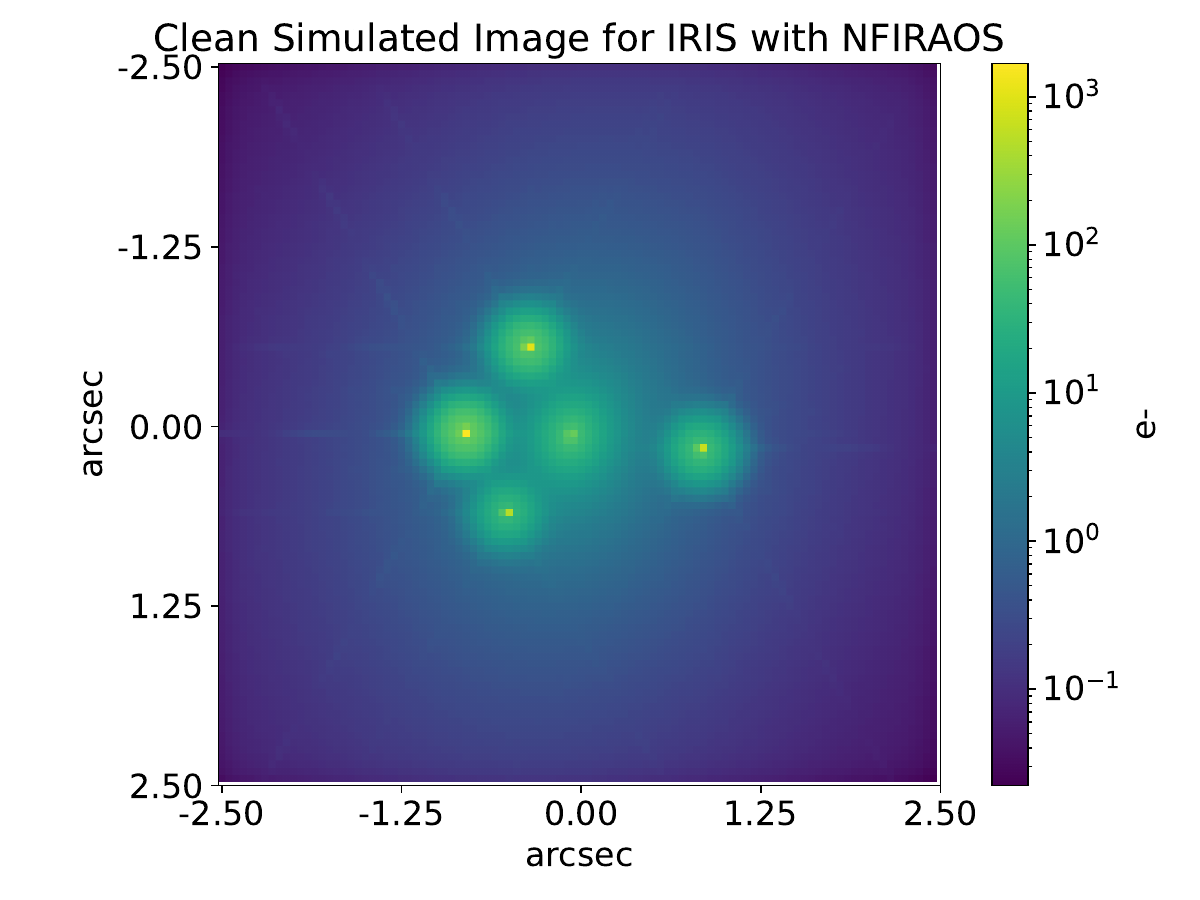} \\
			\hspace{-12mm}
			\includegraphics[width=0.3\textwidth]{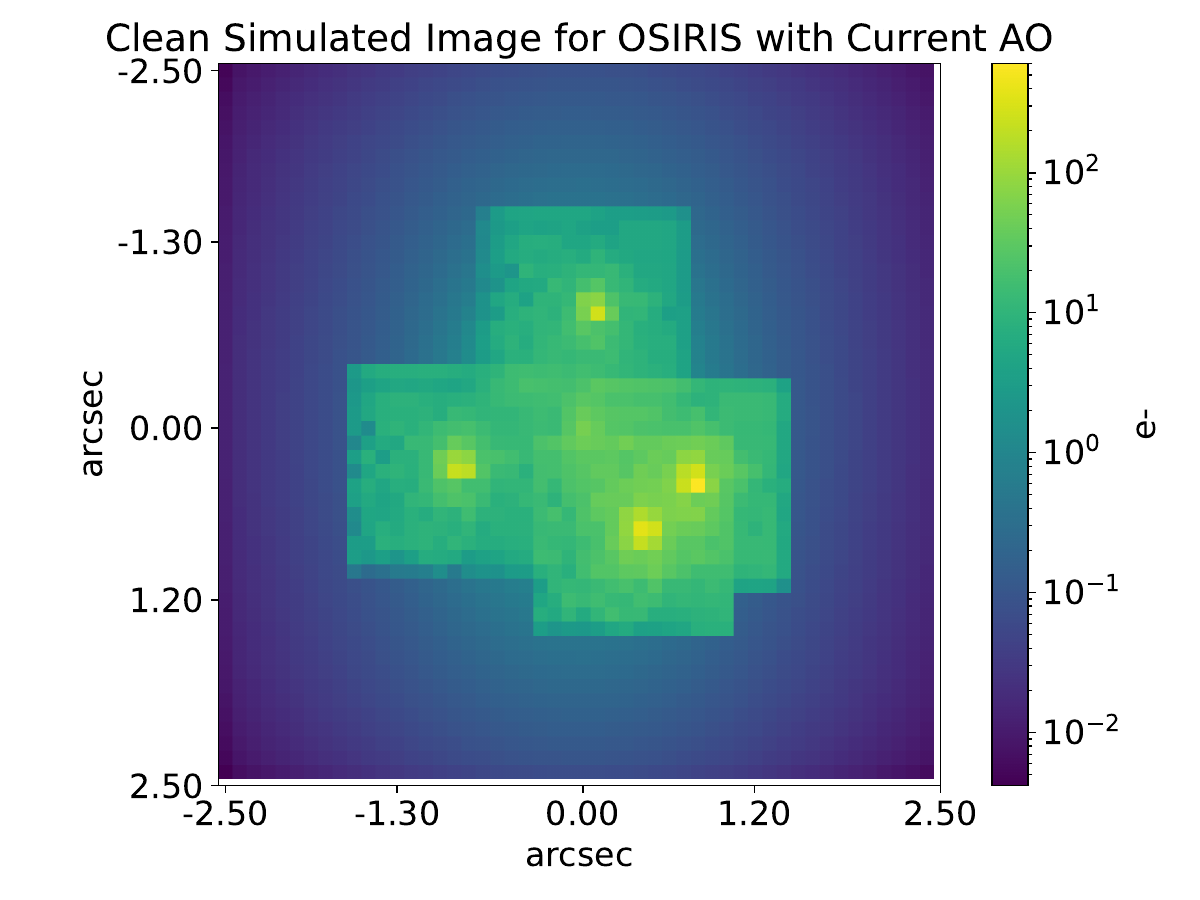} &
			\hspace{-9.2mm}
			\includegraphics[width=0.3\textwidth]{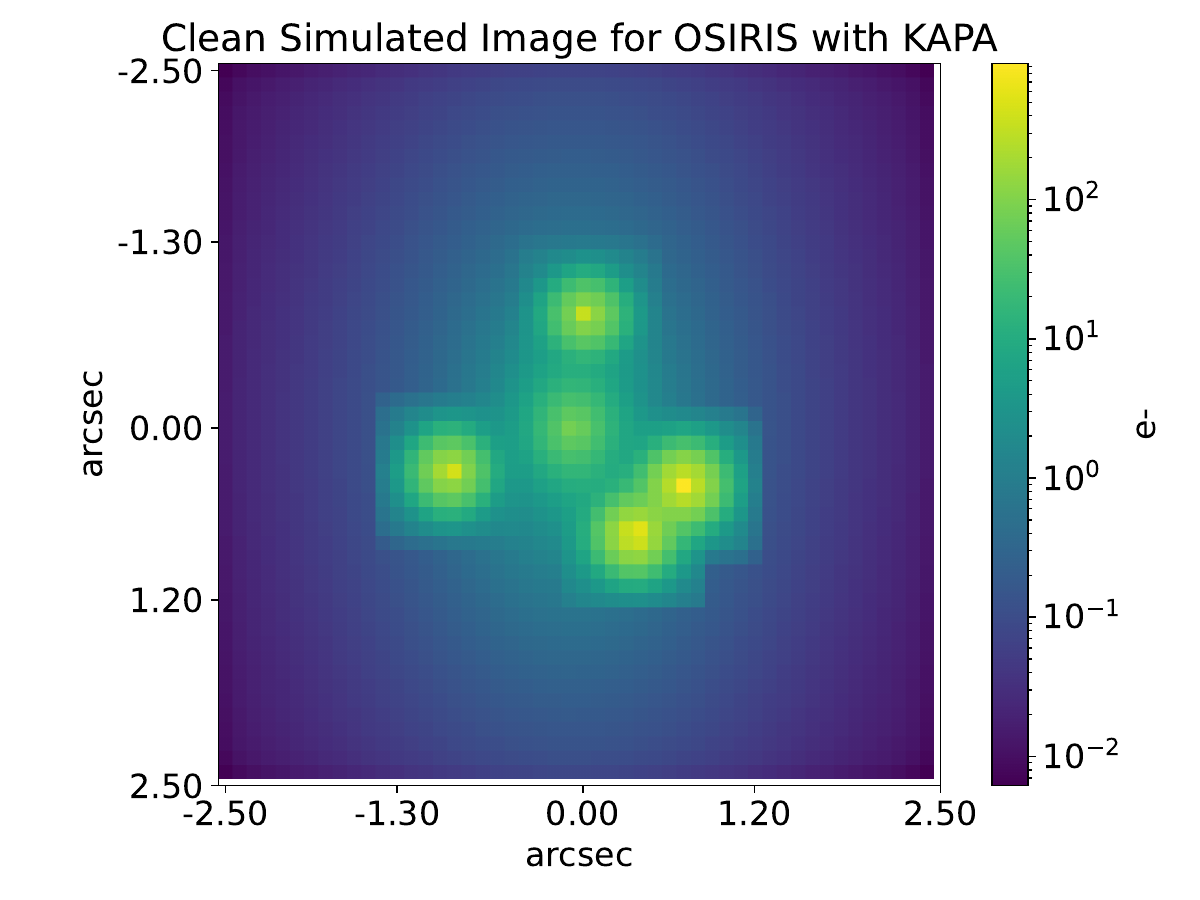} &
			\hspace{-9.2mm}
			\includegraphics[width=0.3\textwidth]{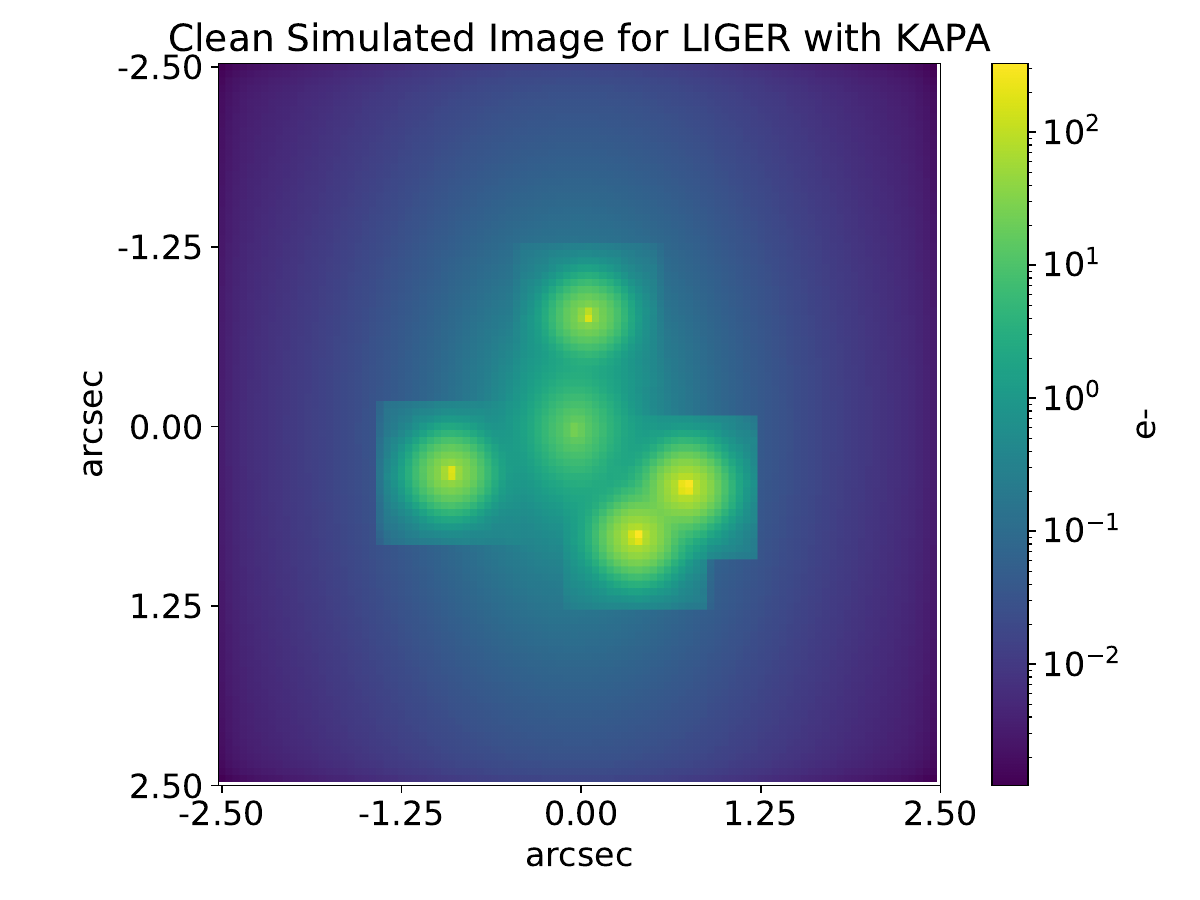} &
			\hspace{-9.2mm}
			\includegraphics[width=0.3\textwidth]{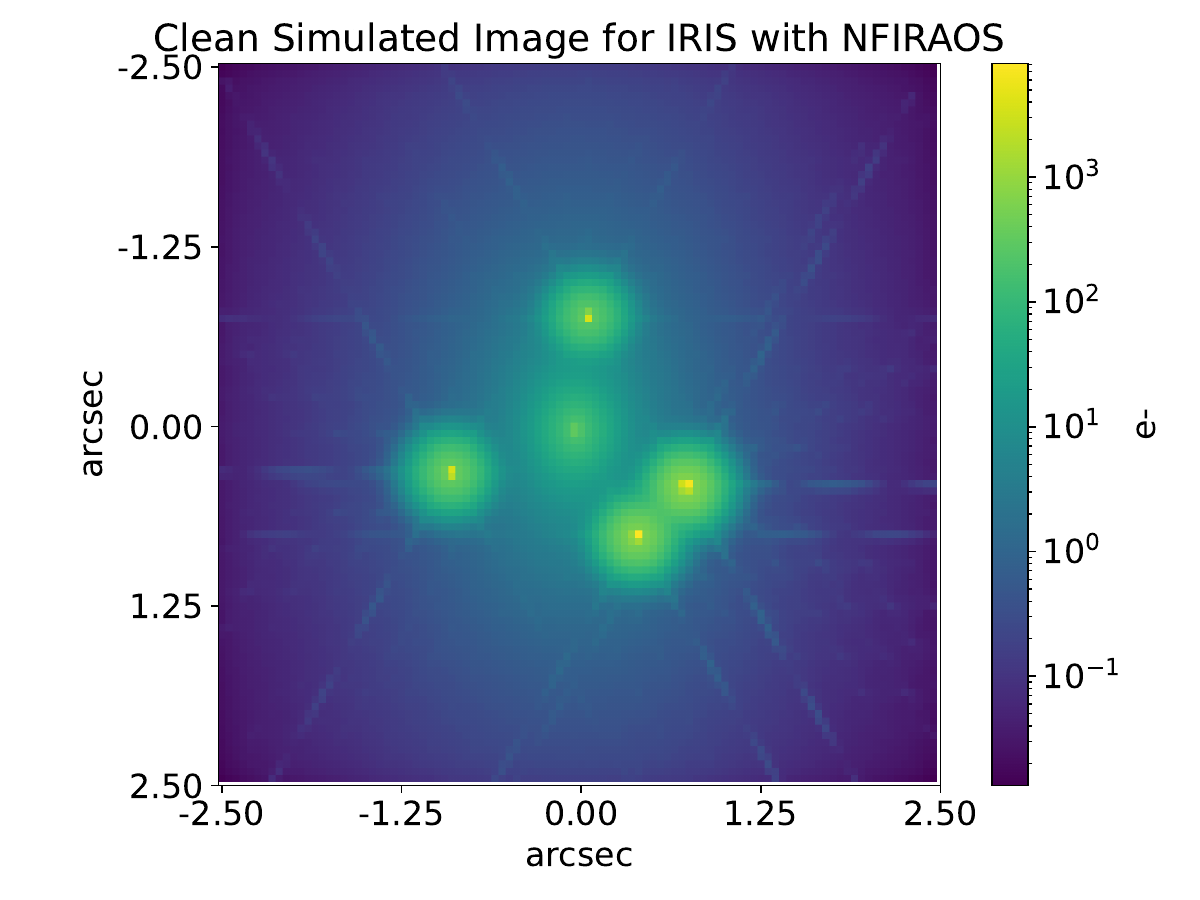} \\
		\end{tabular}
		\caption[]{
        Simulations of a cross (upper row), cusp (middle row) and fold (bottom row) gravitational lens system as they would appear through the 4 different observatory configurations, without noise. From left to right in each row: OSIRIS/current AO, OSIRIS/KAPA, LIGER/KAPA, and IRIS/NFIRAOS. All simulations assume a seeing of 0\farcs3, while the Strehl is 0.3}{ \label{fig:simulated_images_log}}
	\end{center}
\end{figure*}

\begin{figure*}
	\begin{center}
		\begin{tabular}{cccc}
			\hspace{-8mm}
			\includegraphics[width=0.27\textwidth]{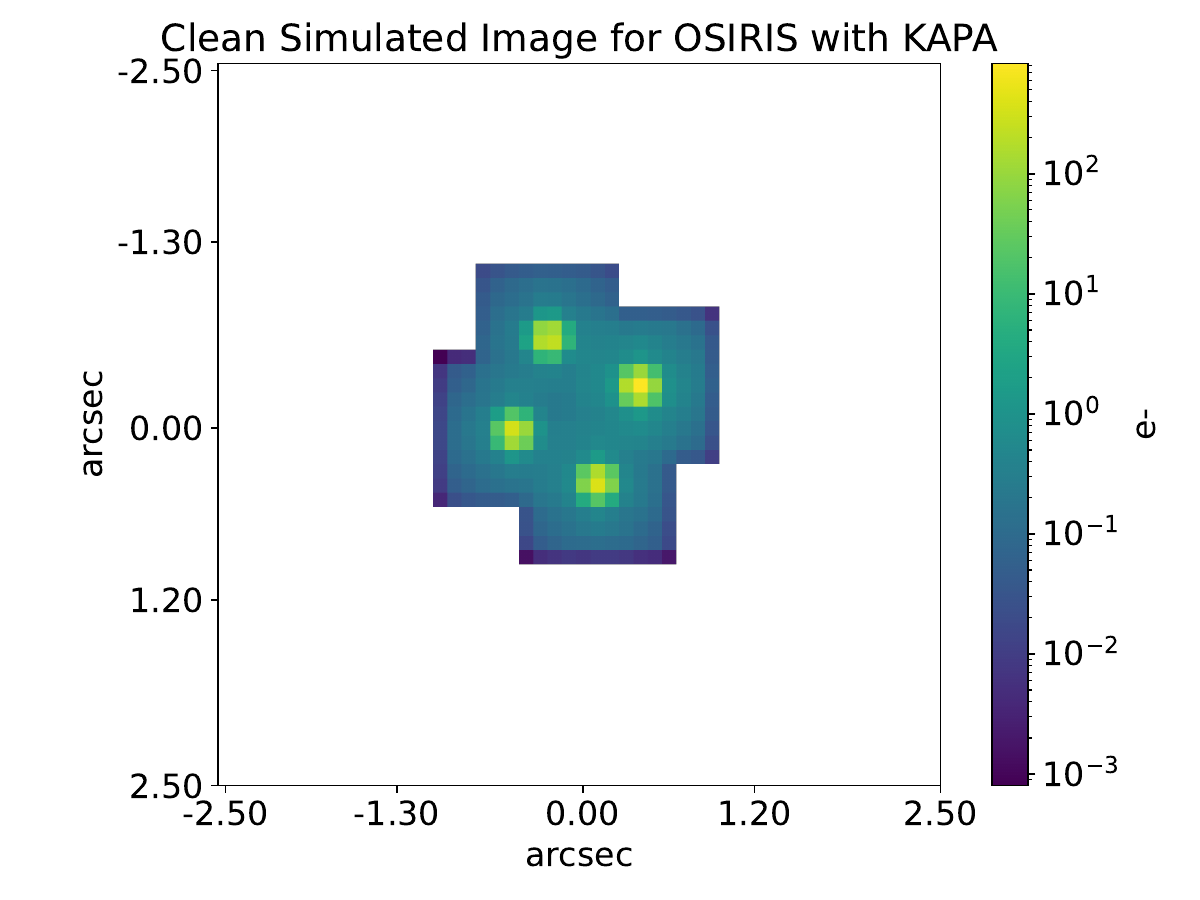} &
			\hspace{-5.2mm}
			\includegraphics[width=0.27\textwidth]{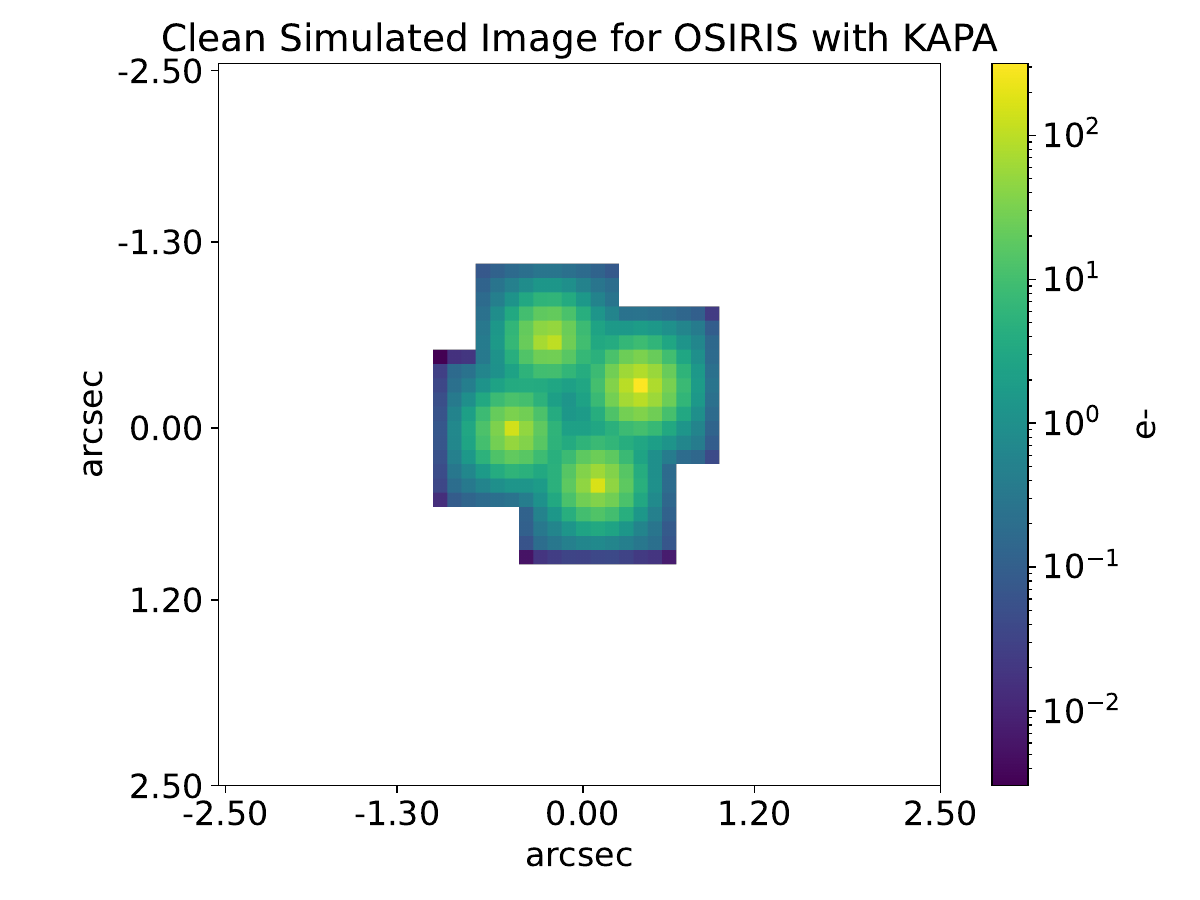} &
			\hspace{-5.2mm}
			\includegraphics[width=0.27\textwidth]{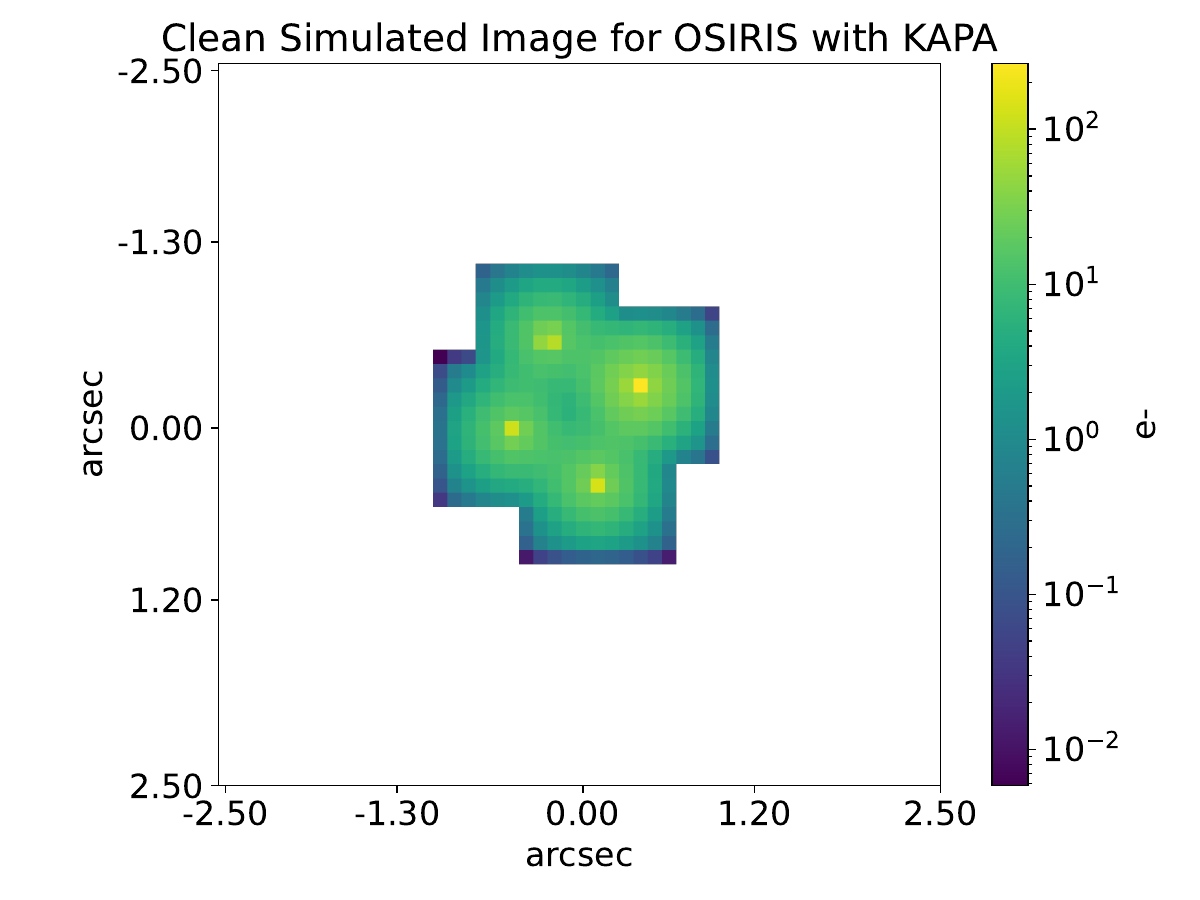} &
			\hspace{-7.2mm}
			\includegraphics[width=0.27\textwidth]{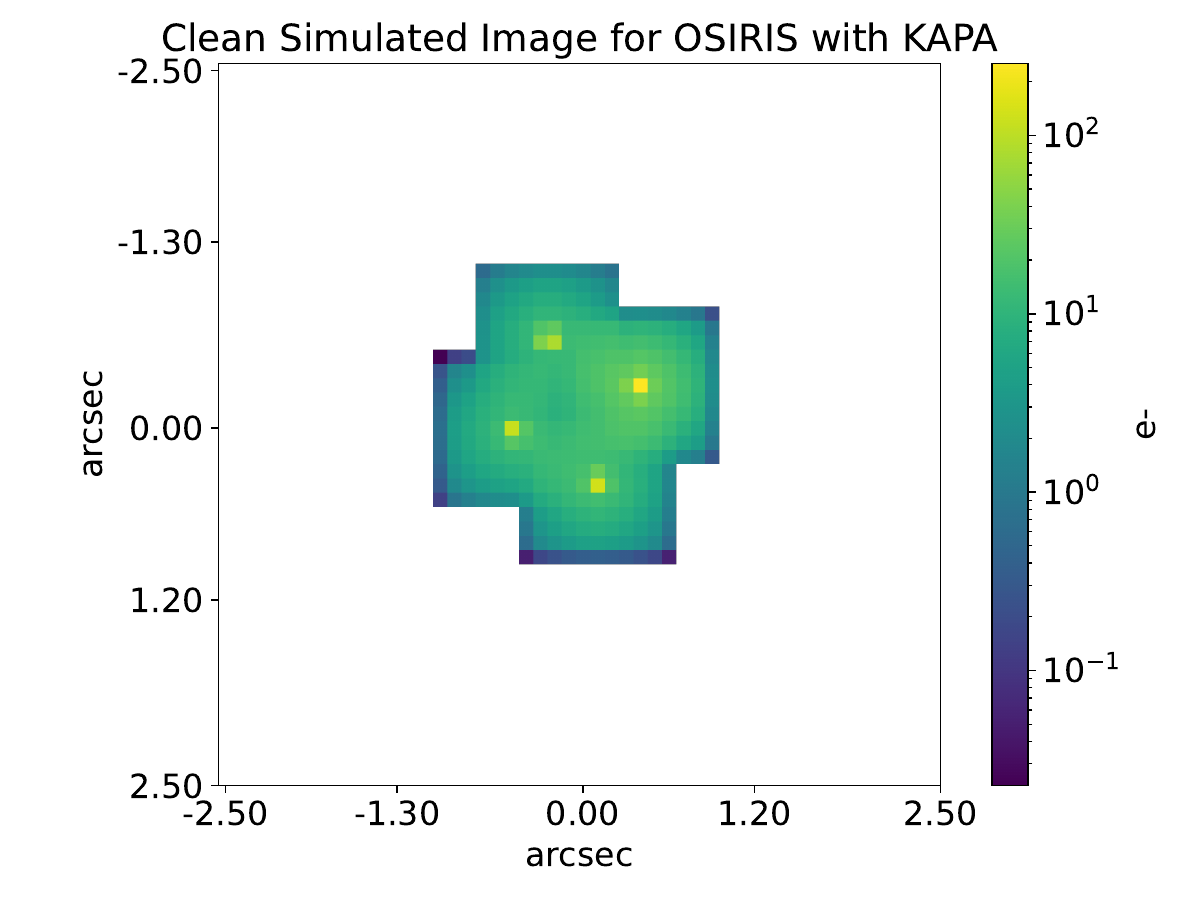} \\
		\end{tabular}
		\caption[]{
  Effect of the seeing: the simulated cross system in the OSIRIS+KAPA configurations for different values of the seeing full-width-half-max: 0.1, 0.3, 0.5, and 0.7 arcsec,from left to right.}{ \label{fig:seeing_images_linear_and_log}}
	\end{center}
\end{figure*}

Fig. \ref{fig:simulated_images_log} shows noiseless simulations of a cross, cusp and fold system as they would appear through the 4 observational configurations: Keck OSIRIS+current adaptive optics system, Keck OSIRIS with KAPA, Keck LIGER with KAPA, and TMT IRIS with NFIRAOS. Fig. 
Figs. \ref{fig:noisy_simulated_images} and Fig. \ref{fig:noisy_simulated_images_log} show the linear and log images with simulated noise.
Fig. \ref{fig:seeing_images_linear_and_log} shows the simulated cross system in the Keck OSIRIS+KAPA configurations for different values of the seeing FWHM.

\section{Inference methodology}\label{sec:inference}

For real measurements of quadruply imaged quasars, the narrow-line flux ratios of the lensed quasar images are extracted in a two-step process, as described by \citet{Nierenberg14}. First, the point spread function, image positions and deflector light (if present) are modelled in an integrated `white-light' image integrated which is a combination of multiple wavelength frames, effectively creating a narrow-band observation. Next, the fluxes of the point sources are extracted as a function of wavelength, and the spectrum for each separate lensed quasar image is modelled to measure the [\ion{O}{3}] fluxes. 

In our analysis, we are interested in the expected improvement in precision due to the higher resolution, sampling, and sensitivity of the next generation of adaptive optics and instruments with respect to the current state of the art. Therefore, we focus our simulated measurements on the white-light PSF and deflector light fitting. We estimate that a fractional improvement in our ability to measure these quantities relative to a current white light image will be reflected in the final expected improvement in the narrow-line flux ratio measurements. In absolute terms, while the astrometric uncertainties from the white light are the appropriate quantity to look at, the photometric uncertainties in the while light will be smaller than what is expected for the narrow lines. This is due to the lower signal to noise ratio of the narrow lines in comparison to the white light, and to the additional uncertainties arising from the process of separating the narrow line emission from the continuum and pseudocontinuum and from the wings of broad H$\beta$. Based on current OSIRIS data (Nierenberg et al. 2024, in prep) we estimate that the uncertainties on the narrow line flux ratios are approximately three times larger than those on the white light flux ratios. Therefore, we apply this scaling when expressing our uncertainties in terms of narrow line flux ratios. In reality, this approximation should be considered a conservative estimate of the full improvement expected from IRIS and LIGER as these instruments will have higher resolution spectral sampling which will enable improved sky-line subtraction and spectral decomposition and therefore enhance the spectral fitting part of this measurement.

\begin{figure*}
\begin{center}
\begin{tabular}{cccc}
    \hspace{-5.20mm}
	\includegraphics[width=0.5\textwidth]{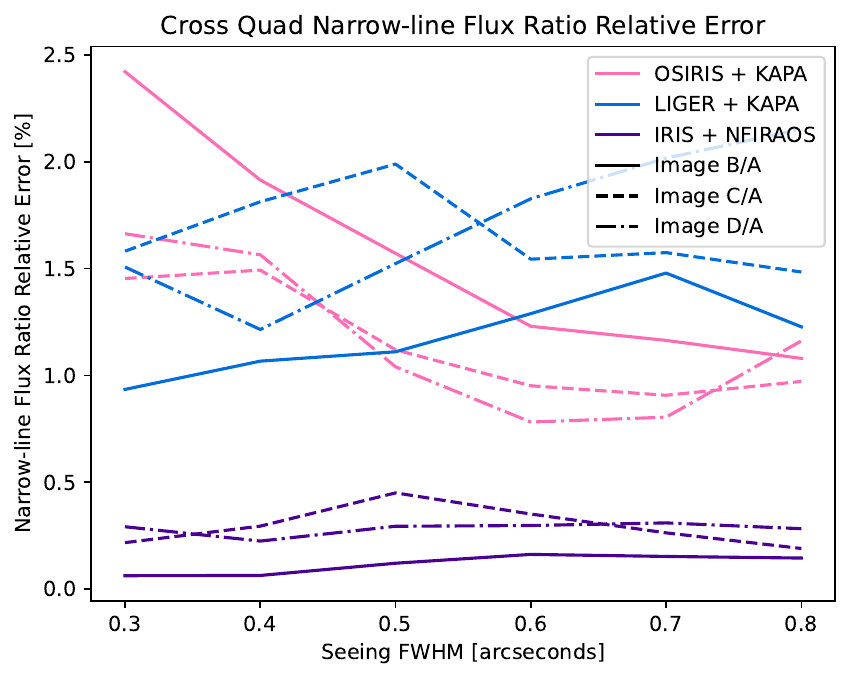}
 &
    \hspace{-4.70mm}
	\includegraphics[width=0.5\textwidth]{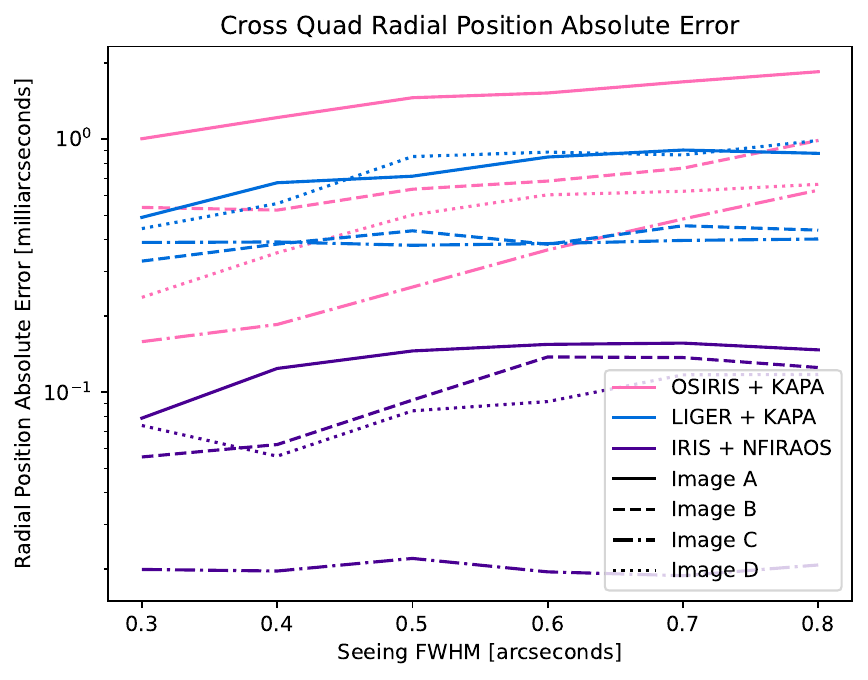}\\
    \hspace{-5.20mm}
	\includegraphics[width=0.5\textwidth]{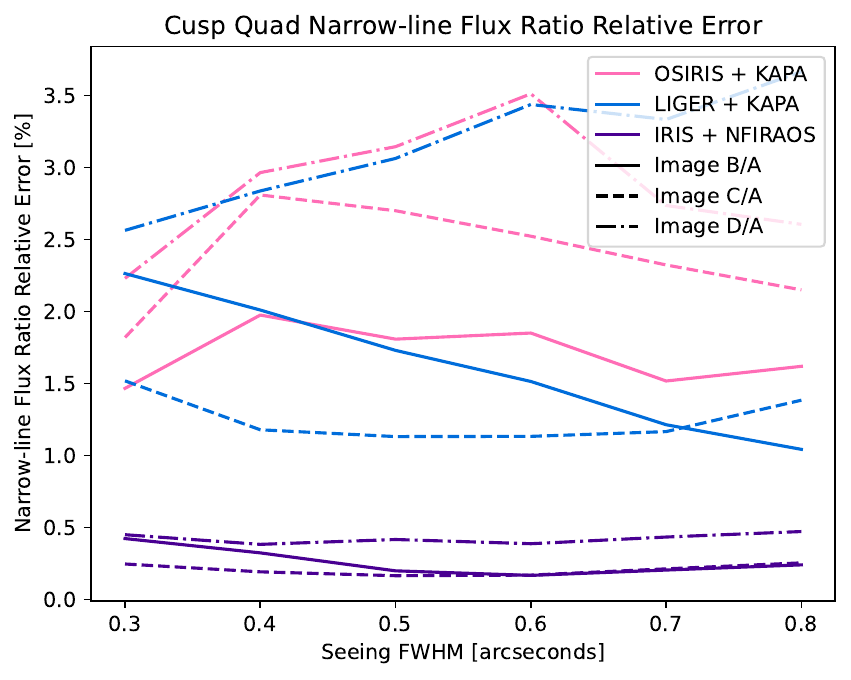}
 &
    \hspace{-4.70mm}
	\includegraphics[width=0.5\textwidth]{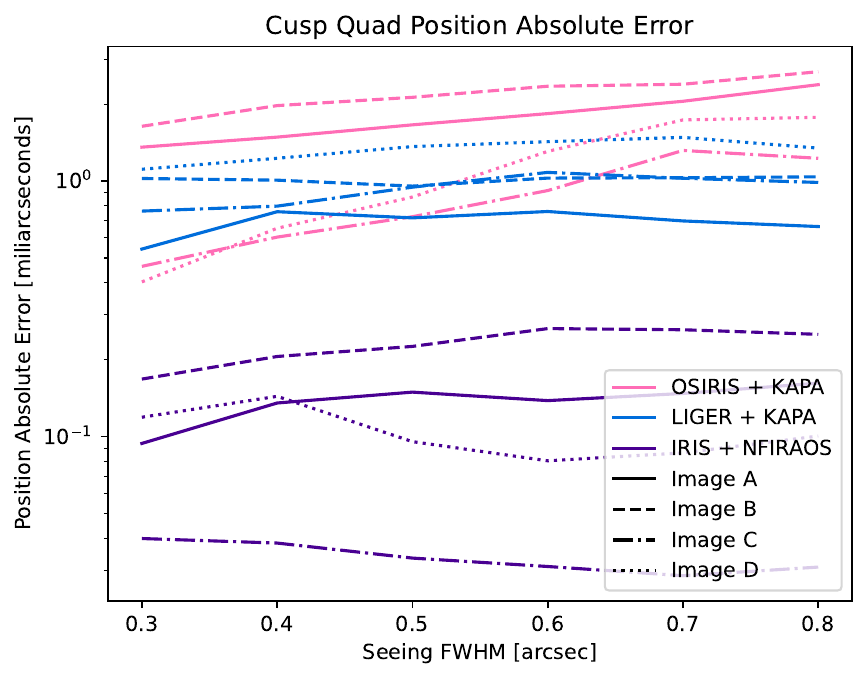}\\
    \hspace{-5.20mm}
	\includegraphics[width=0.5\textwidth]{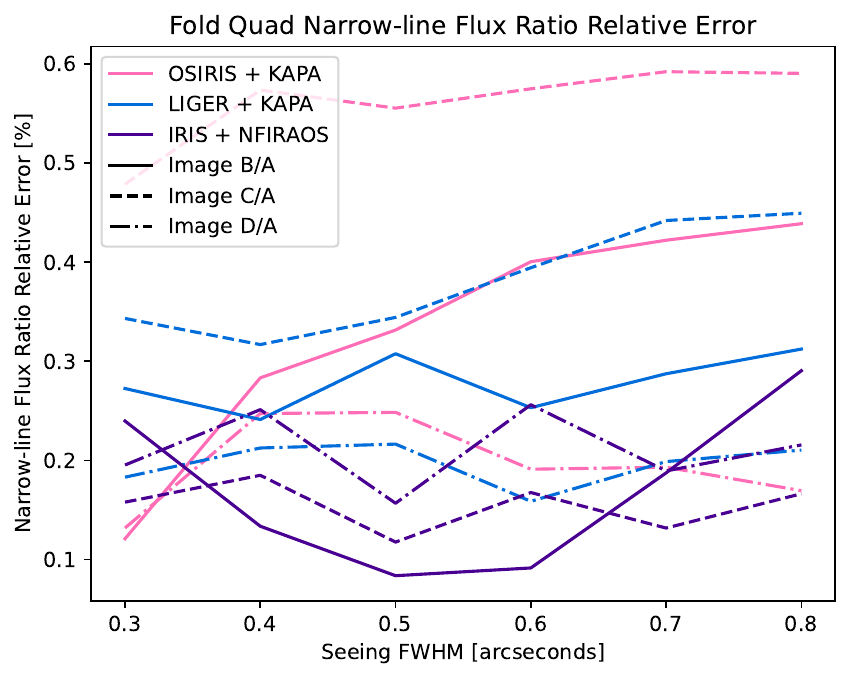}
 &
    \hspace{-4.70mm}
	\includegraphics[width=0.5\textwidth]{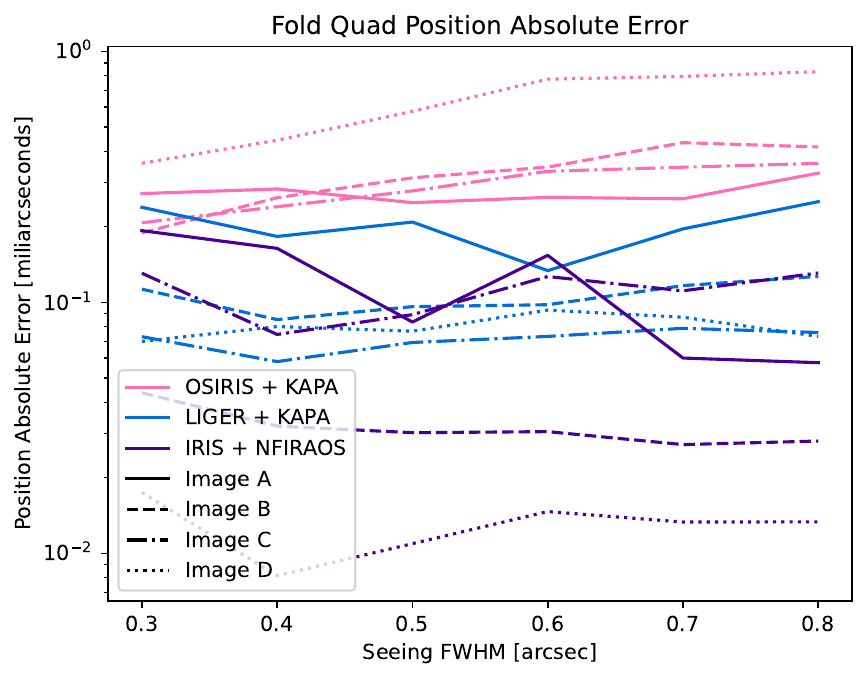}
\end{tabular}
\caption[]{ 
Left panels:  the flux ratio error percentage as a function of seeing, for OSIRIS+KAPA, LIGER+KAPA, and IRIS+NFIRAOS system configurations overlayed on the same plot, for the cross, cusp and fold system from top to bottom. The flux ratios are calculate for images B, C, D relative to image A. The flux ratio percentile represent the ratio between  the standard deviation of the MCMC samples and mean  of the samples, expressed  as a percentile.  Right panels: the error in the astrometry of the quasar images positions, for the same system configurations as the left panels, obtained from the standard deviation of the MCMC samples. We see that the flux error percentiles are solidly under the target of $2\%$, which each system having better and better precision. These simulations were done under an assumption of a known PSF with a strehl of 0.3 (which is lower than the expected strehl values for these systems.) }{ \label{fig:master_known_psf_error_analysis}}
\end{center}
\end{figure*}

\begin{figure*}
	\begin{center}
		\begin{tabular}{cccc}
			\hspace{-5.20mm}
			\includegraphics[width=0.5\textwidth]
            {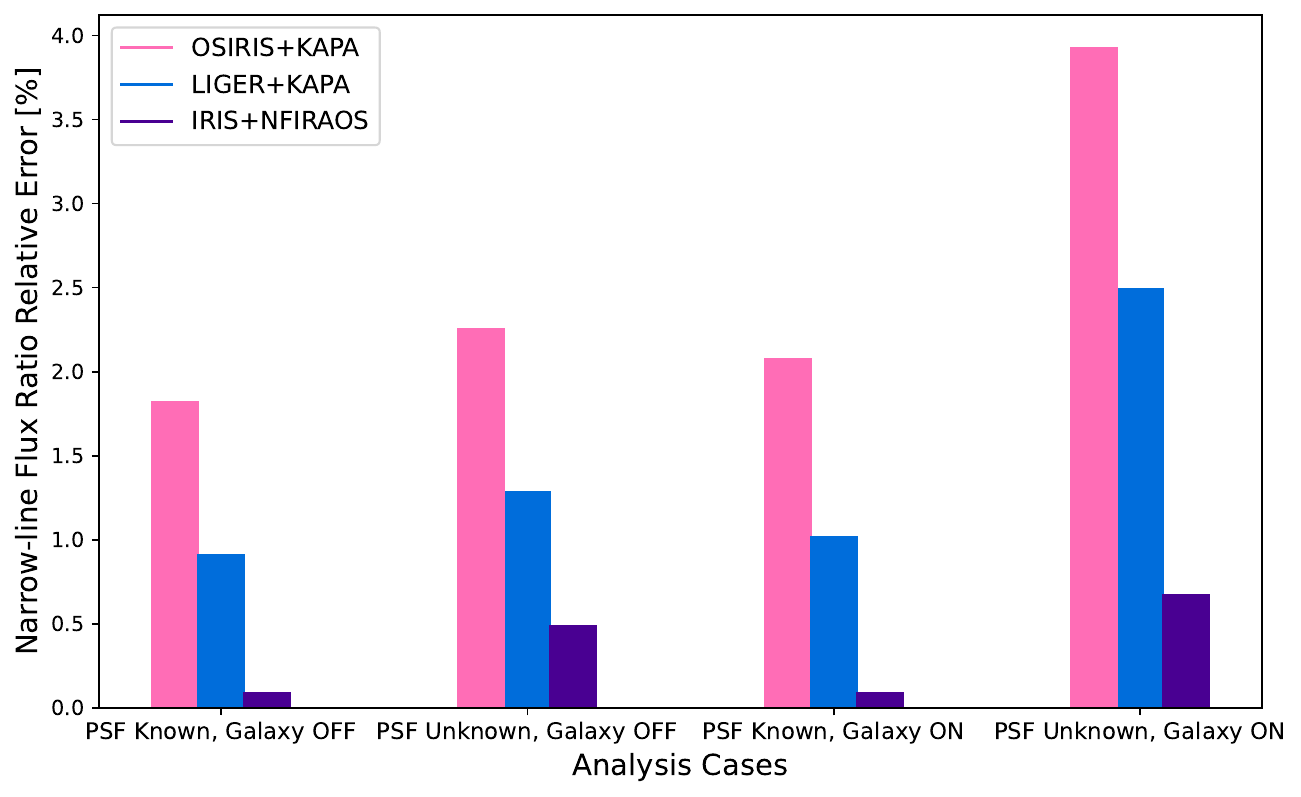}
			&
			\hspace{-4.70mm}
			\includegraphics[width=0.5\textwidth]{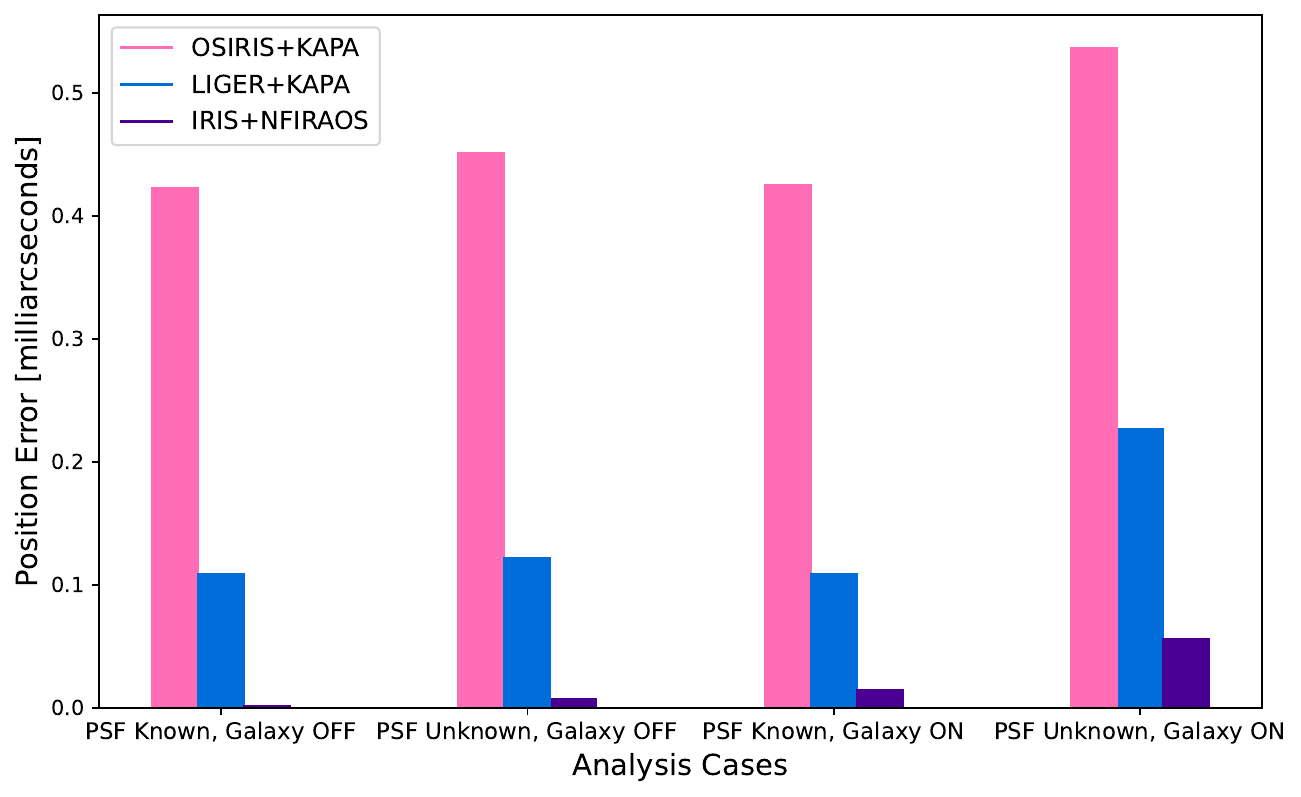}
		\end{tabular}
		\caption[]{Comparison of the astrometric and photometric performance for each observational setup, with known and unknown PSF and with or without the deflector surface brightness (ON or OFF). At the wavelength of the narrow emission lines, the deflector galaxy will typically be negligible. We include the case with the deflector galaxy surface brightness included to illustrate the performance in case of particularly weak quasar emission lines. The 2\% and mas precision targets are virtually met in all cases, except for OSIRIS with unknown PSF and a bright deflector galaxy (ON). For TMT IRIS+NFIRAOS the statistical uncertainties are negligible for the simulated 600s observations, implying that a few minutes of observations will be sufficient for typical lensed quasars. The simulations shown here assumed aseeing of $0\farcs3$ and Strehl of 0.5.}{ \label{fig:bar_plot}}
	\end{center}
\end{figure*}

Our inference pipeline has the following steps: 
\begin{itemize}[noitemsep,topsep=0pt,partopsep=0pt]
    \item Simulate the quad images with specific detector, AO system PSF, seeing and Strehl.
    \item Add noise corresponding to the specific detector and exposure time.
    \item Initialize the model parameters that will be used for the fitting, including the lensed images, the lens galaxy and the PSF.
    \item Run a particle swarm optimizer (PSO) to find the best fitting model parameters.
    \item Use the results of the PSO to inform the initialization points for the Markov Chain Monte Carlo method used to sample the posterior for the parameters.
\end{itemize}

The PSF can either be passed as a known array to model, or inferred by the software through multiple PSF iterations before starting the PSO optimization, exploiting the built-in information provided by the four quasar images. For generality, we consider two cases. One in which the surface brightness of the main deflector galaxy is included and one in which it is excluded. The choices are meant to bracket reality. There are systems in which at the wavelength of interest, the deflector will not be bright enough to warrant modeling, while in certain systems it will be prominent and will need to be modeled. 

\section{Results}\label{sec:results}

Figure~\ref{fig:master_known_psf_error_analysis} compares the flux ratio and astrometric uncertainties as a function of seeing FWHM across the four lens configurations.

The relative errors in flux are below the threshold target of $2\%$.  Similarly, astrometric precision is reliably within the milliarcsecond range, meeting the overall target. 
We note that these simulations are rather conservative, assuming a Strehl ratio of 0.3, which should be generally exceeded by future AO systems.
Furthermore, both astrometric and photometric precision improve substantially going from the current system to the upcoming ones.
Thus, with future facilities it will be possible to extract more information per lens and thus achieve higher sensitivity, e.g., to warm dark matter models than currently possible \citep{gilman2019b}.


In order to consider the effects of two additional non negligible sources of error, in Figure \ref{fig:bar_plot} we compare how the measurement precision is affected by the inclusion of the deflector light (i.e. when the narrow emission lines are comparable to the foreground continuum) and whether the PSF is known prior to the inference  as a result or modeling the AO system, or whether it needs to be inferred from the data (denoted as 'ON' or 'OFF').   





Again, the target precision is met virtually in every case, except for the one of unknown PSF and bright lens galaxy for OSIRIS + KAPA. Even in the most conservative case of unknown PSF, the statistical uncertainties obtained with TMT IRIS+NFIRAOS are very small for our adopted 600s exposures, indicating that much shorter exposure times will be sufficient, thus greatly enhancing survey speed. For example, in order to obtain the same uncertainties obtained with LIGER+KAPA in 600s, 84s are sufficient with IRIS+NFIRAOS.

Appendix Figs.~\ref{fig:flux_posterior} and \ref{fig:position_posterior} show the posteriors obtained from one the MCMC sampling of the OSIRIS+KAPA case. The posteriors are Gaussian, and the original flux ratios are recovered well.

We note that our results can easily be scaled to predict the performance of similar instrumentation on the Giant Magellan Telescope (GMT). Since the mirror diameter is the main driver of the both astrometric and photometric precision, AO corrected instruments on GMT will achieve precision between Keck LIGER+KAPA configurations and TMT IRIS+NFIRAOS systems.

%

\section{Conclusions}

We simulated quadruply imaged quasars to estimated the performance in terms of astrometry and narrow line flux ratios of current and upcoming state-of-the-art adaptive optics+instrument system for ground telescopes. As an illustration we have considered the Keck Telescope( OSIRIS + KAPA, LIGER+KAPA), and the Thirty Meter Telescope (IRIS+NFIRAOS). 
We have shown that expected errors in flux ratio will be well below $2\%$, and that milliarcsecond precision will be achievable in terms of astrometry, even if the PSF needs to be reconstructed from the data. 

With TMT IRIS+NFIRAOS exposure times of 1-2 minutes will be sufficient for the lenses simulated here, implying that it will be practical to characterize the large numbers of lensed quasars that will soon be discovered by the Euclid, Rubin, and Roman Telescopes, including the majority that will be significantly fainter than the ones known so far. 

\section*{Acknowledgements}


IZ and TT acknowledge support by the National Science Foundation grant NSF-1836016, ``Astrophysics enabled by Keck All Sky Precision Adaptive Optics",
by the Gordon and Betty Moore Foundation Grant 8548 ``Cosmology via Strongly lensed quasars with KAPA'', 
and by NASA grant HST-GO-15177. Part of the data used in this paper were obtained as part of HST-GO-15177.

TT and AMN acknowledge support by the NSF through grant "Collaborative Research: Measuring the physical properties of darkmatter with strong gravitational lensing".

IZ acknowledges this work was supported by the Natural Sciences and Engineering Research
Council of Canada (NSERC), [funding reference \#DIS-2022-568580].

We thank Prof. Shelley Wright and Dr. Nils Rundquist for providing PSF configurations for KAPA and NFIRAOS, and for helpful discussions.

This research made use of the \textit{lenstronomy} package \cite{Birrer18} to model the quad lensed systems, and the following software: the NASA Astrophysics Data System's Bibliographic Services (ADS),  the color blindness palette by Martin Krzywinski and Jonathan Corum\footnote{\url{http://mkweb.bcgsc.ca/biovis2012/color-blindness-palette.png}}, the Color Vision Deficiency PDF Viewer by Marie Chatfield \footnote{\url{https://mariechatfield.com/simple-pdf-viewer/}}, 
Jupyter Notebook \citep{Kluyver2016jupyter}, Matplotlib \citep{Hunter2007}, NumPy \citep{VanderWalt2011}, Python \citep{Millman2011, Oliphant2007}, scikit-learn \citep{Pedregosa2012}. Open AI's GPT4 was used to check for grammatical correctness of the phrasing.

\section*{Data Availability}

The Hubble Space Telescope data used as a model for the quad lensed systems simulated in this study can be found at the HST archive.
%


\bibliographystyle{mnras}
\bibliography{Adaptive_Optics.bib,more-refs,references-big-22}



\appendix

\section{Noise estimation}\label{sec:app_noise}

\paragraph*{Current Adaptive Optics System}
For the current AO system, we use the noise levels inferred from a real life observation done at Keck in the Hbb band, for an exposure time of 600s.
For the LIGER and IRIS noise estimation, we use the same formula as \cite{meng2015}:
\begin{equation}
\text{Var}_{\text{pix}} = C \cdot t +B\cdot t +N_{\text{read}} R^2
\end{equation}

where $C$ is the signal from clean lens system in electrons per second per pixel, $t$ is the integration
time in seconds, $B$ is the sum of the sky background and detector dark current in electrons per second
per pixel, $N_{\text{read}}$ is the number of detector readouts, and R is the standard deviation of the read noise
in electrons. All the properties pertaining to each detector are shown in Table \ref{table:instrument_specs}.

\begin{align*}
\text{sky\_frame\_value} &= b \left[ (S+DC) \cdot t + N \cdot R^2 \right]^{0.5} \\
\text{dark\_frame\_value} &= b \left( DC \cdot t + N \cdot R^2 \right)^{0.5}
\end{align*}

\begin{equation}
S = \frac{1}{t} \left( \left( \frac{\text{sky\_frame\_value}}{\text{dark\_frame\_value}} \right)^2 \left( DC \cdot t + N_{\text{read}} \cdot R^2 \right) - N_{\text{read}} \cdot R^2 \right) - DC
\end{equation}
To infer the background sky component for LIGER and IRIS, we use the instrumental properties and the measured noise for the current AO system to scale the noise level accordingly.



\section{Modeling of the real quad system data}
\subsection{Lens light model parameters }\label{sec:lens_conversion}

Conversion between the lens galaxy light model parameters shown in \cite{schmidt2023} and the parameters used by the $lenstronomy$ software package that was used for lens modeling.

\begin{equation}
e_1 =  \frac{1 - q_{L}}{1 + q_{L}} \cos(2\phi_{L})
\end{equation}
\begin{equation}
e_2 =  \frac{1 - q_{L}}{1 + q_{L}} \sin(2\phi_{L})
\end{equation}
$amp$ is equal to $I$, $n_{\text{sersic}}$ is the same, $R_{\text{sersic}} = \theta_E$.




\begin{figure*}
	\begin{center}
		\begin{tabular}{cccc}
			\hspace{-12mm}
			\includegraphics[width=0.3\textwidth]{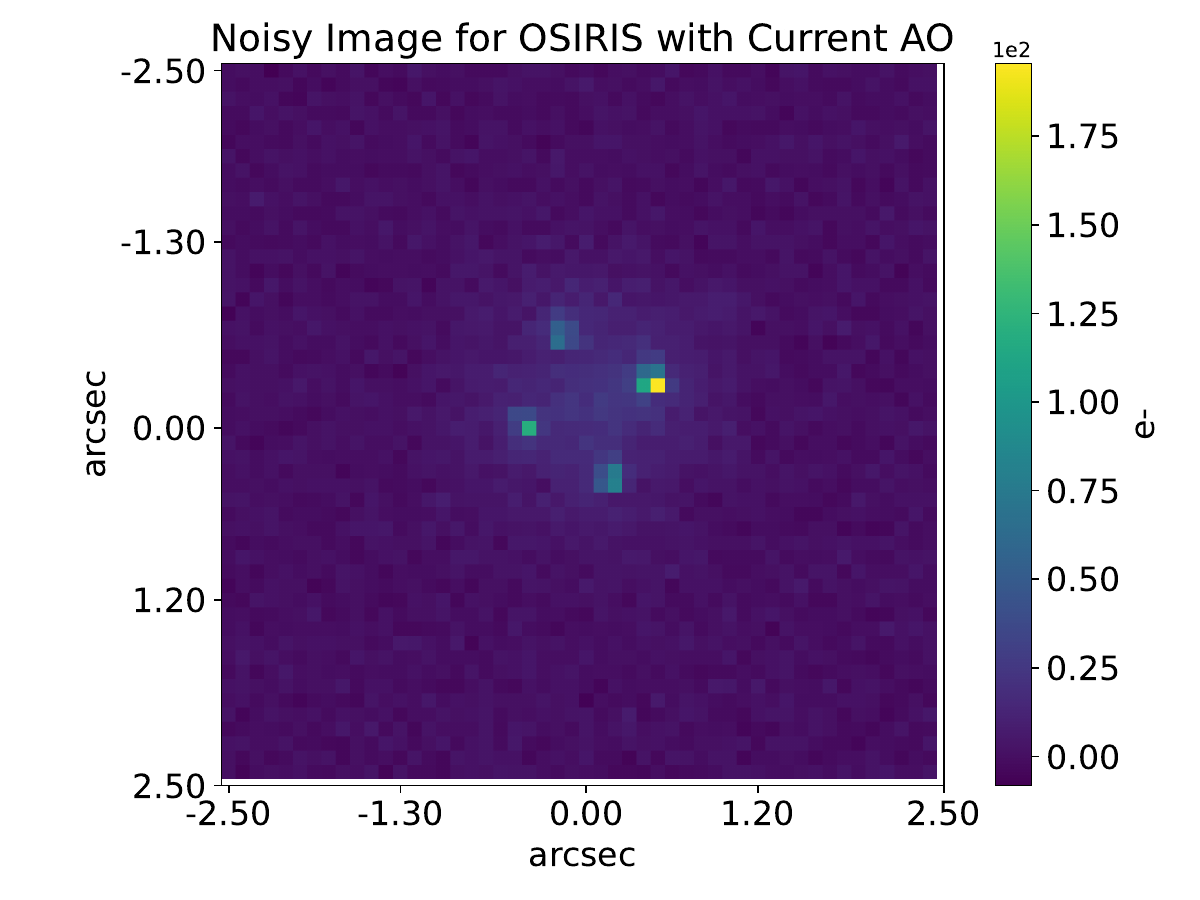} &
			\hspace{-9.2mm}
			\includegraphics[width=0.3\textwidth]{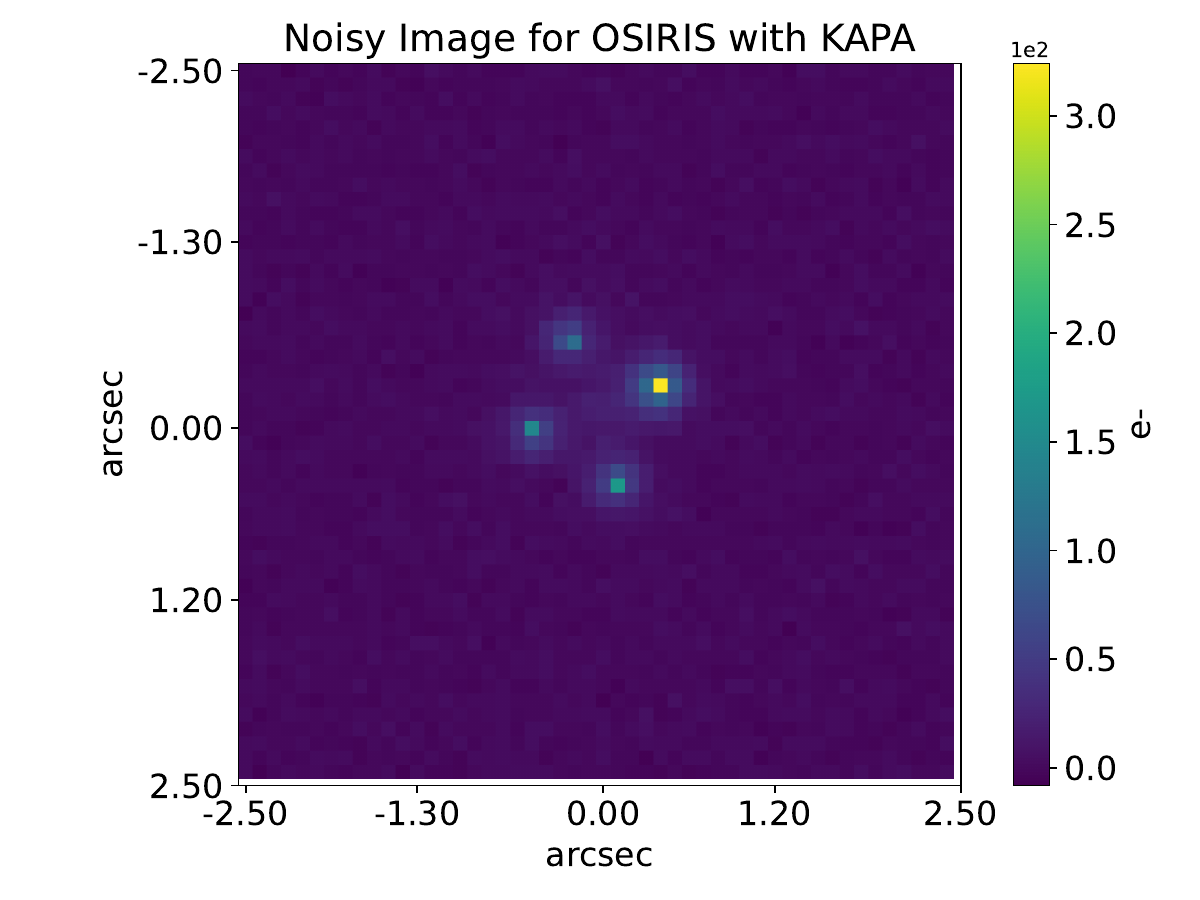} &
			\hspace{-9.2mm}
			\includegraphics[width=0.3\textwidth]{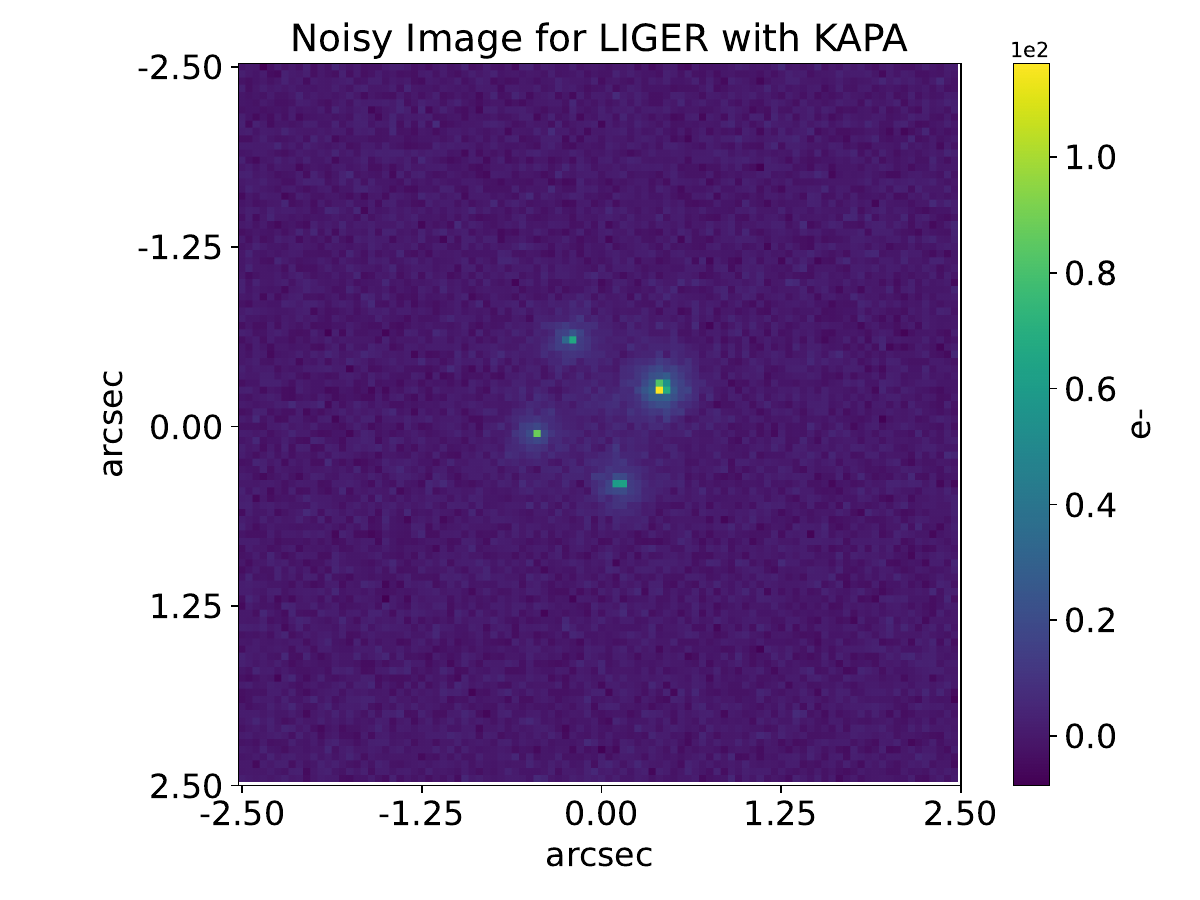} &
			\hspace{-9.2mm}
			\includegraphics[width=0.3\textwidth]{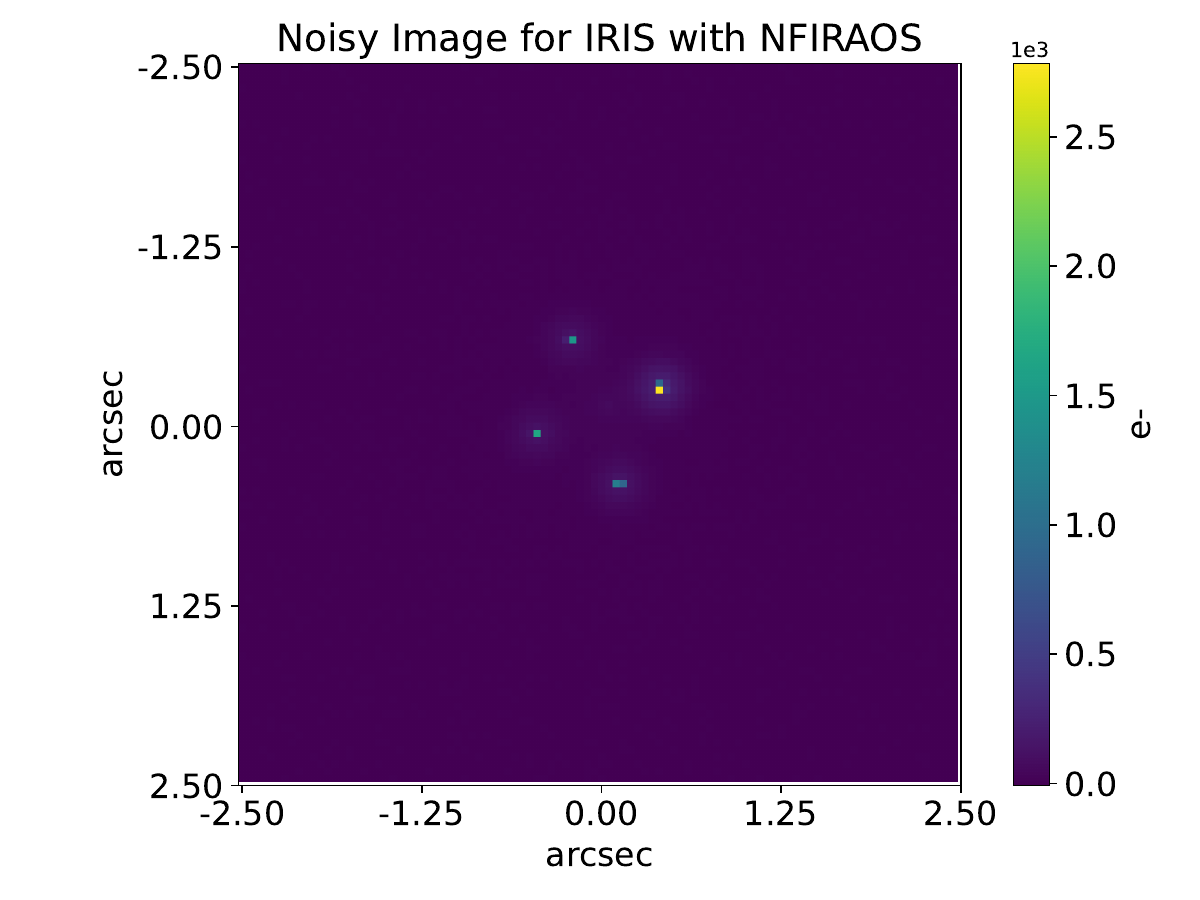} \\
			\hspace{-12mm}
			\includegraphics[width=0.3\textwidth]{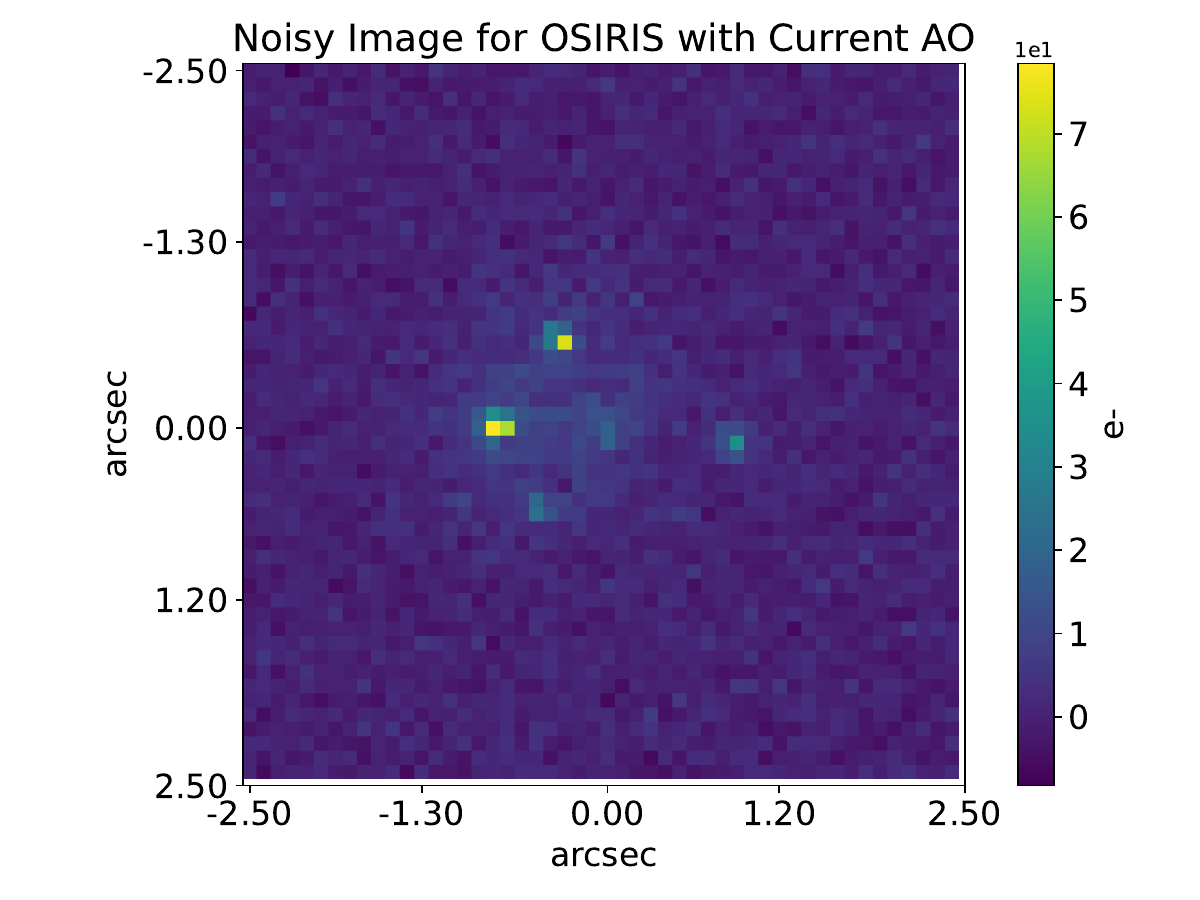} &
			\hspace{-9.2mm}
			\includegraphics[width=0.3\textwidth]{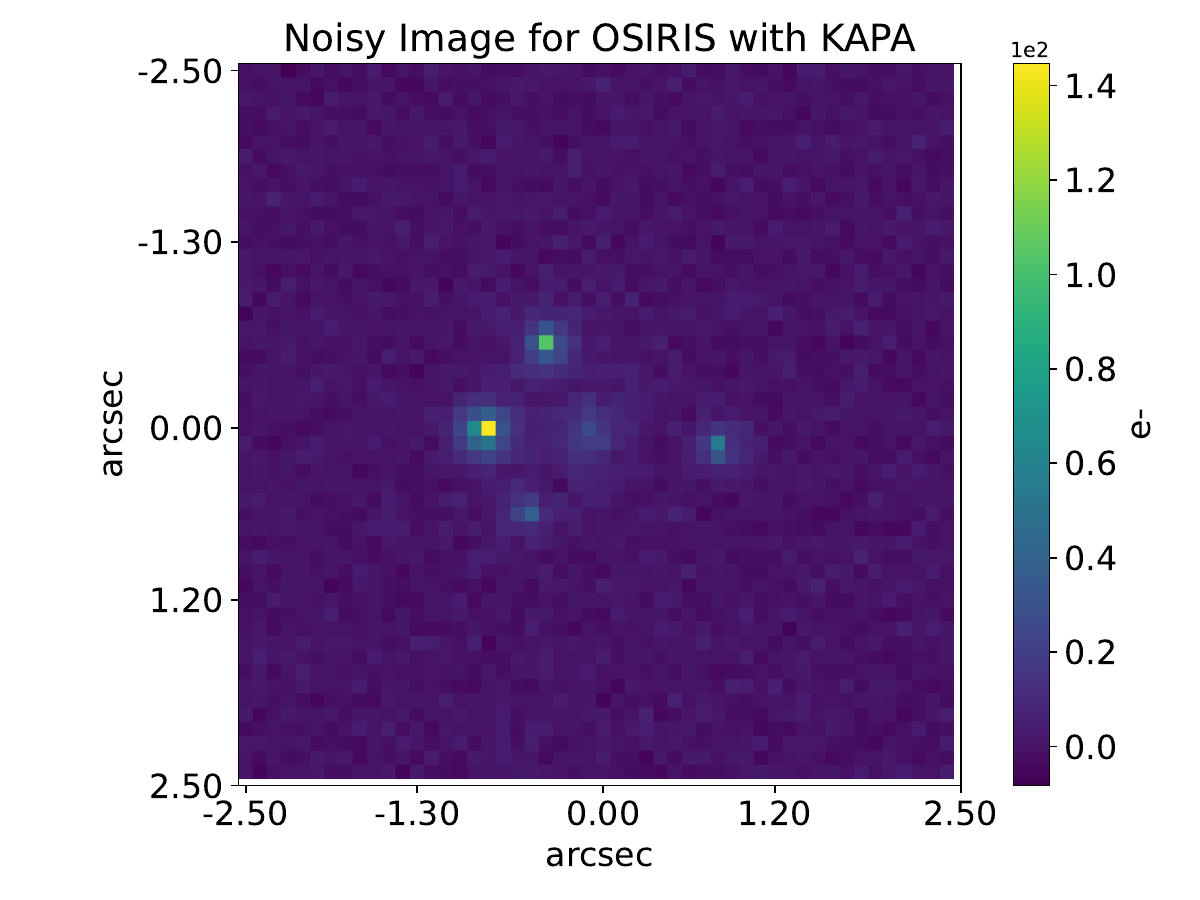} &
			\hspace{-9.2mm}
			\includegraphics[width=0.3\textwidth]{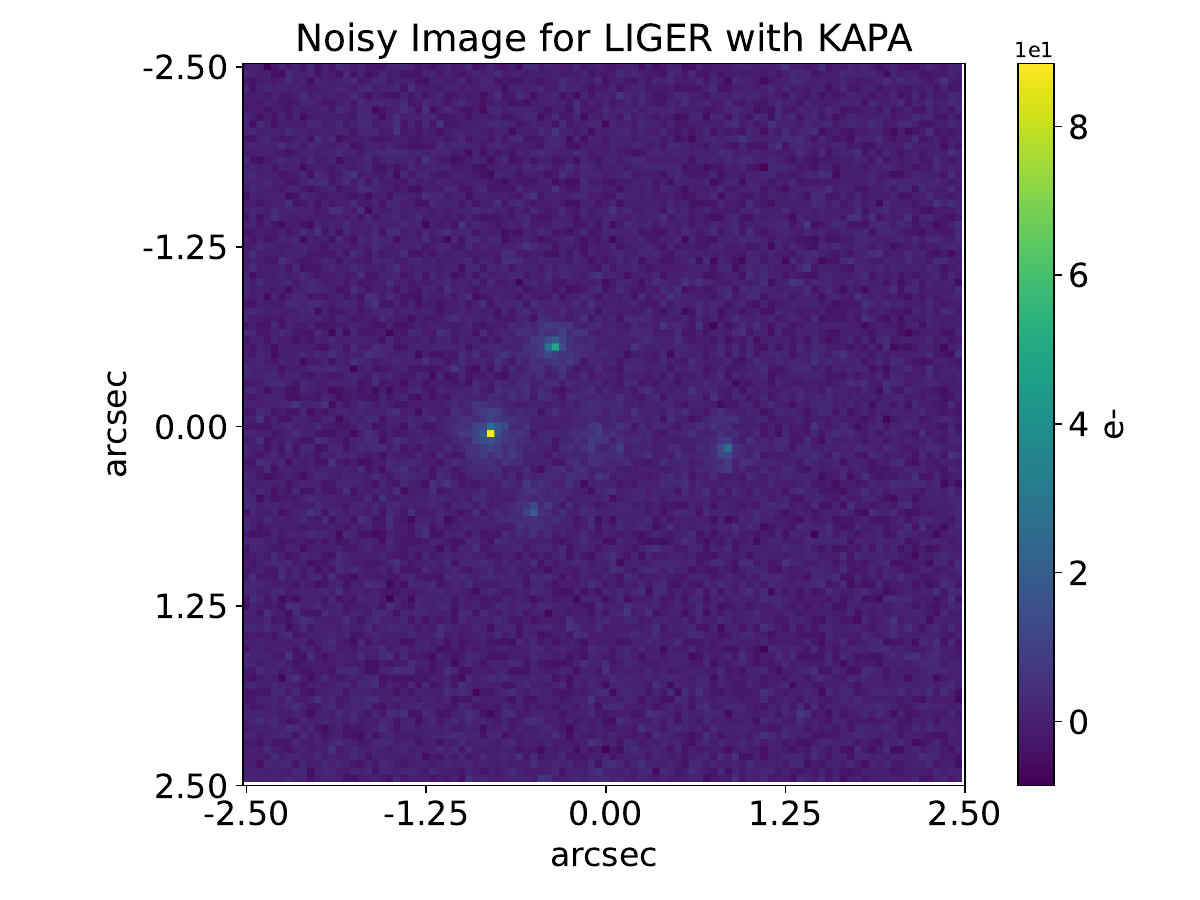} &
			\hspace{-9.2mm}
			\includegraphics[width=0.3\textwidth]{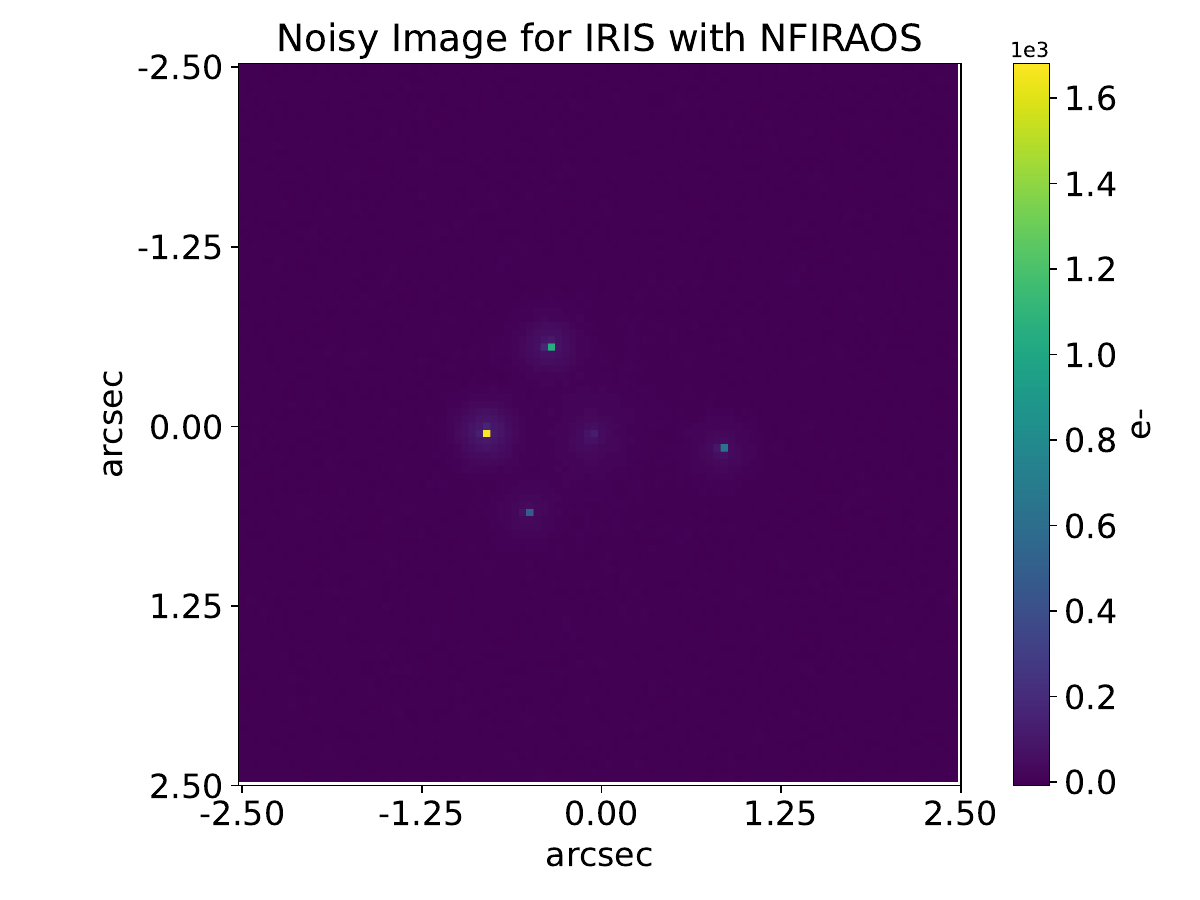} \\
			\hspace{-12mm}
			\includegraphics[width=0.3\textwidth]{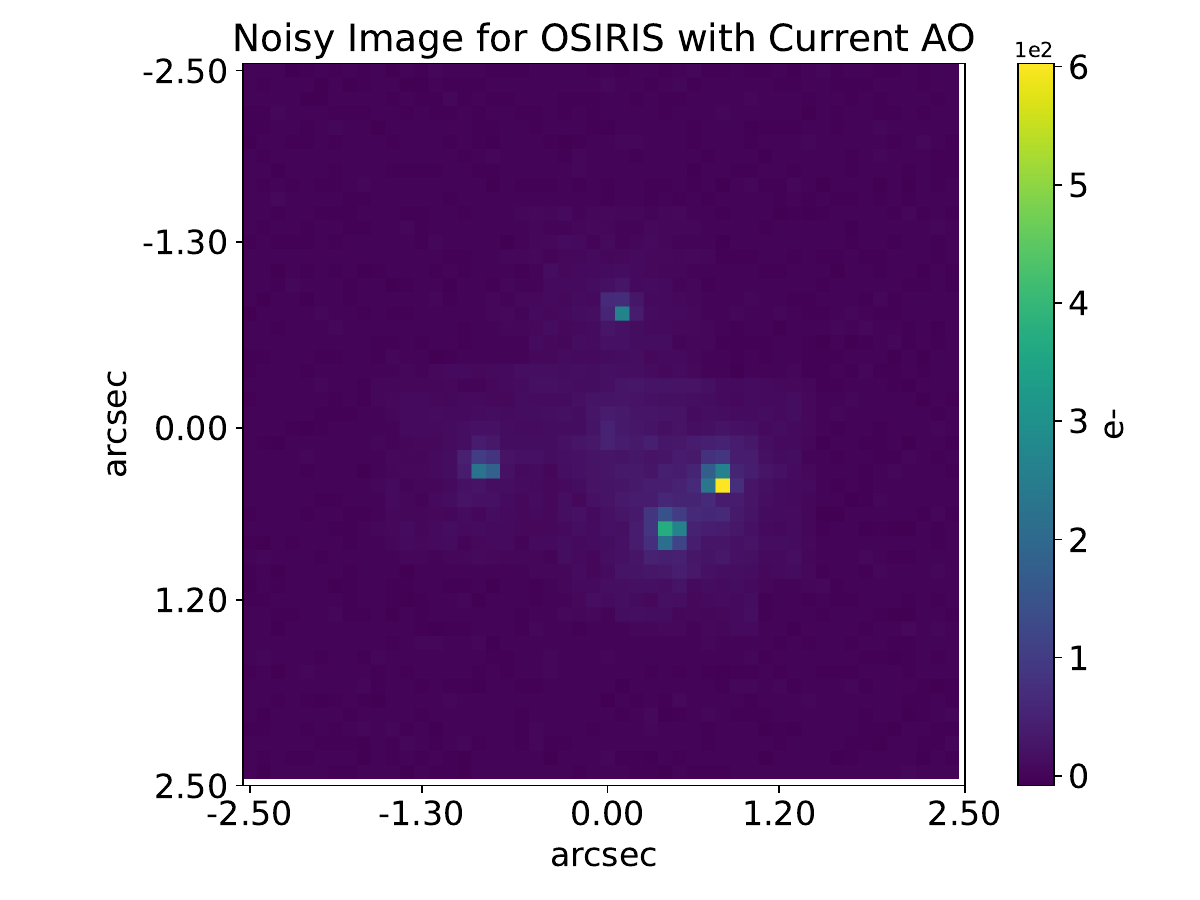} &
			\hspace{-9.2mm}
			\includegraphics[width=0.3\textwidth]{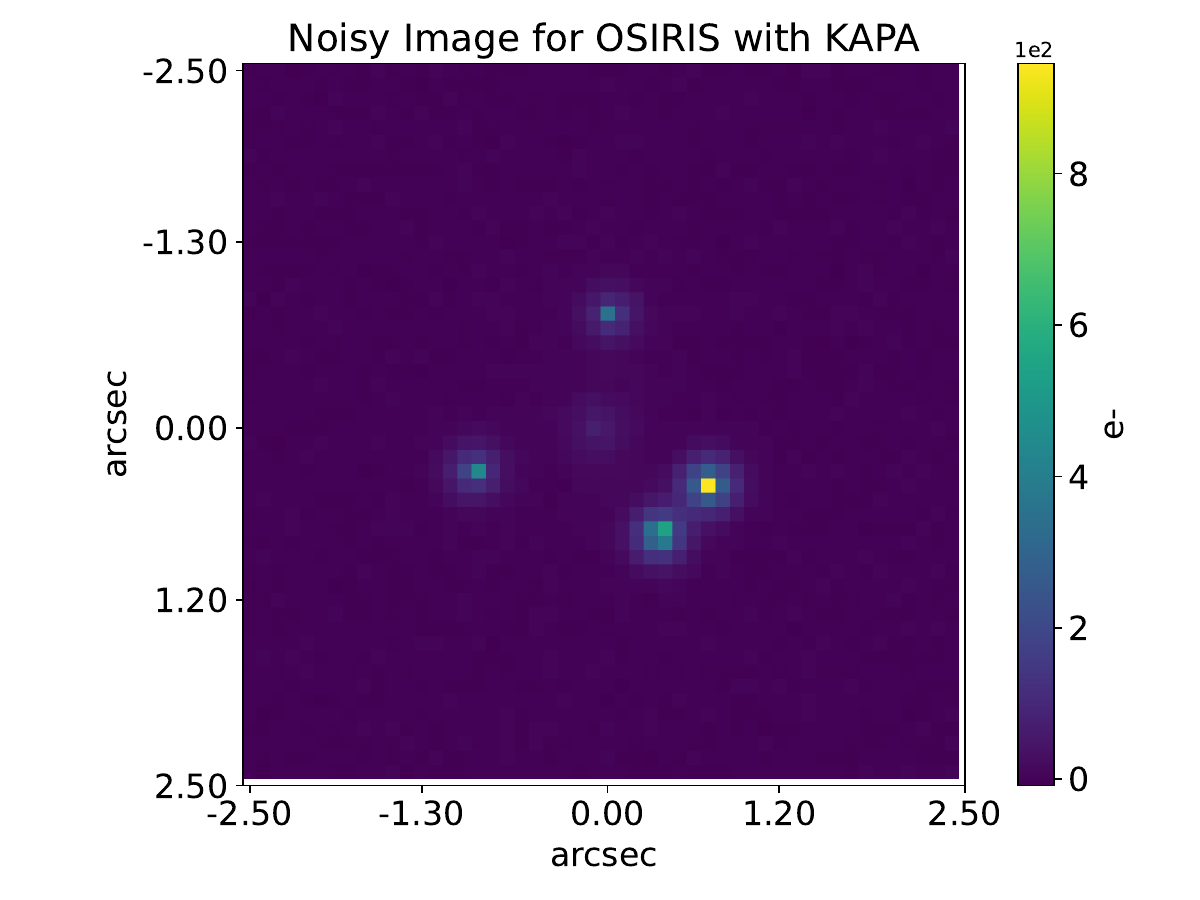} &
			\hspace{-9.2mm}
			\includegraphics[width=0.3\textwidth]{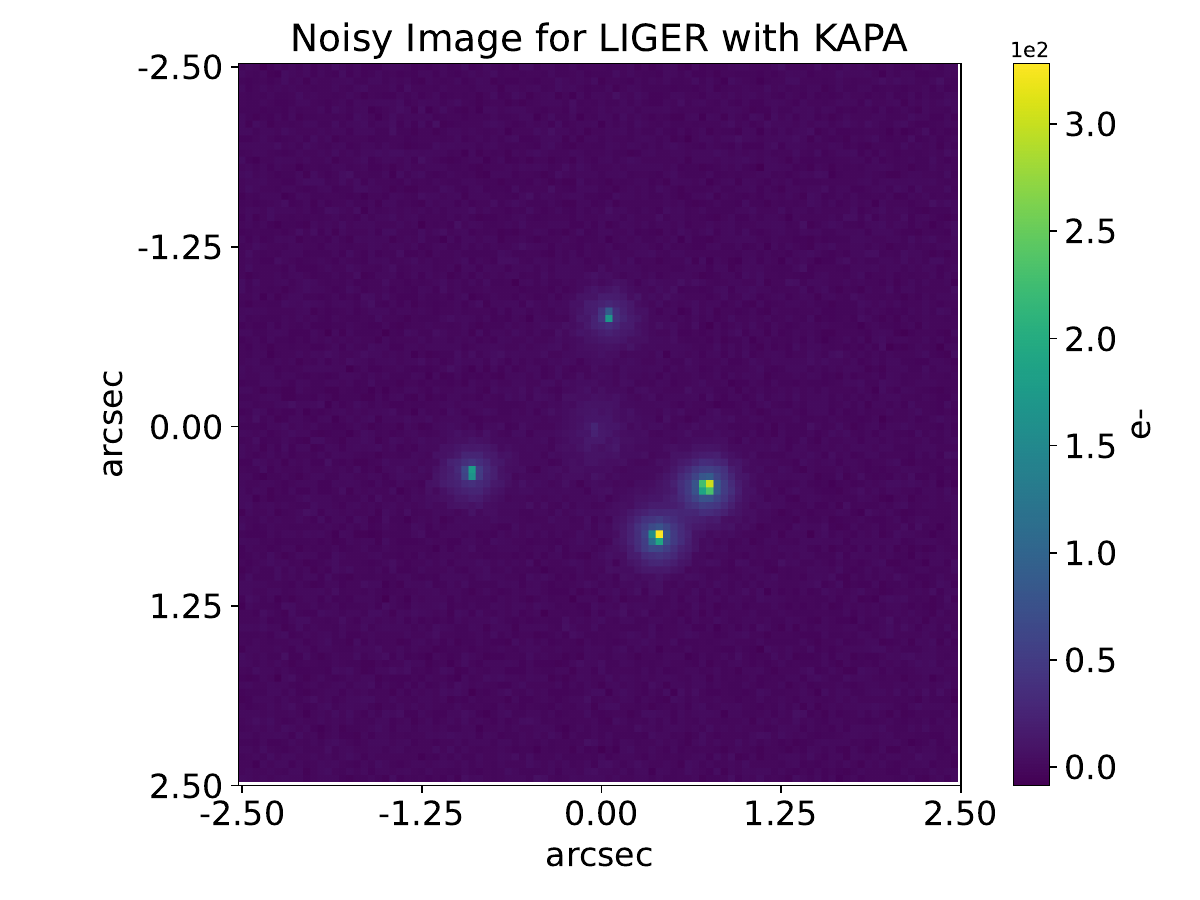} &
			\hspace{-9.2mm}
			\includegraphics[width=0.3\textwidth]{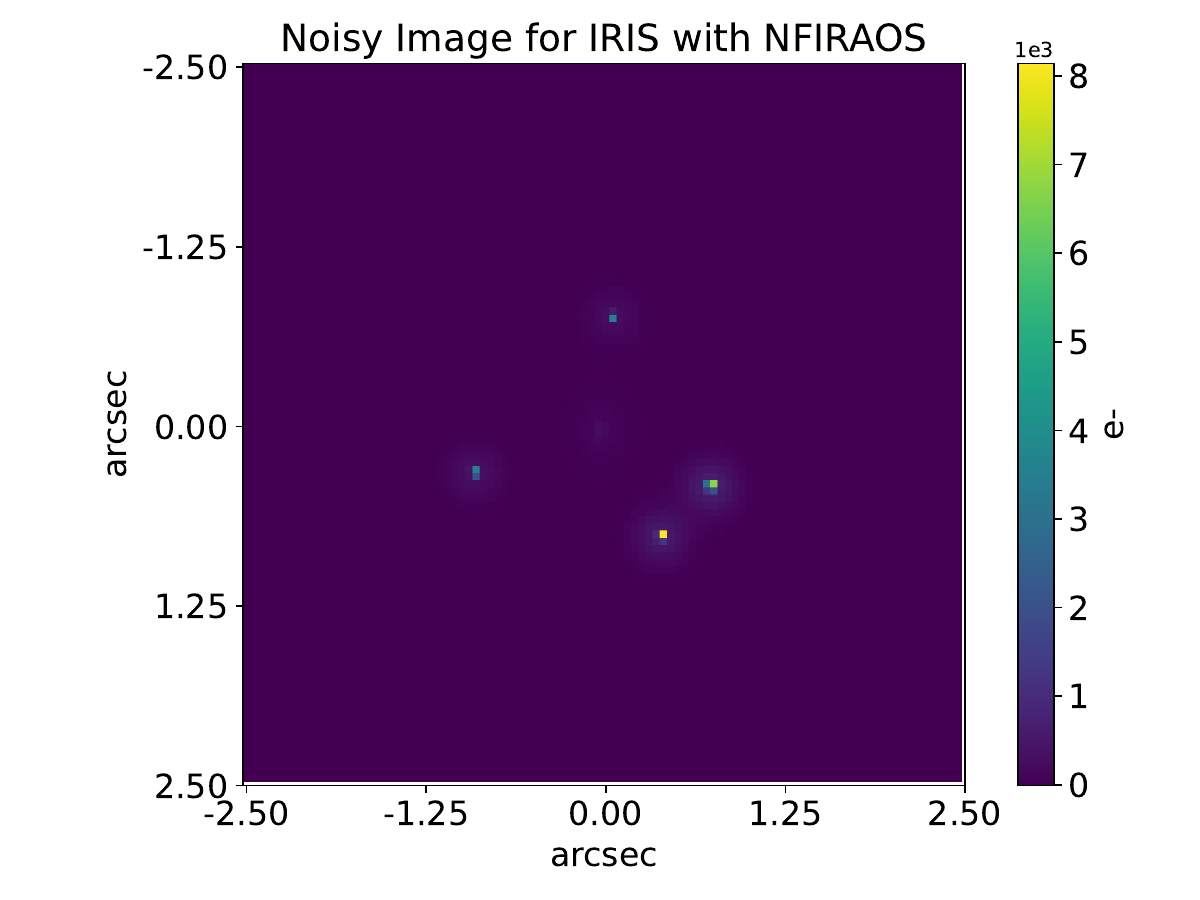} \\
		\end{tabular}
		\caption[]{ Simulations of a cross, cusp and fold system as they would appear through the 4 different observatory configurations: OSIRIS+current adaptive optics system, OSIRIS with KAPA, LIGER with KAPA, and IRIS with NFIRAOS, with the noise levels corresponding to each system overlayed as well.}{ \label{fig:noisy_simulated_images}}
	\end{center}
\end{figure*}

\begin{figure*}
	\begin{center}
		\begin{tabular}{cccc}
			\hspace{-12mm}
			\includegraphics[width=0.3\textwidth]{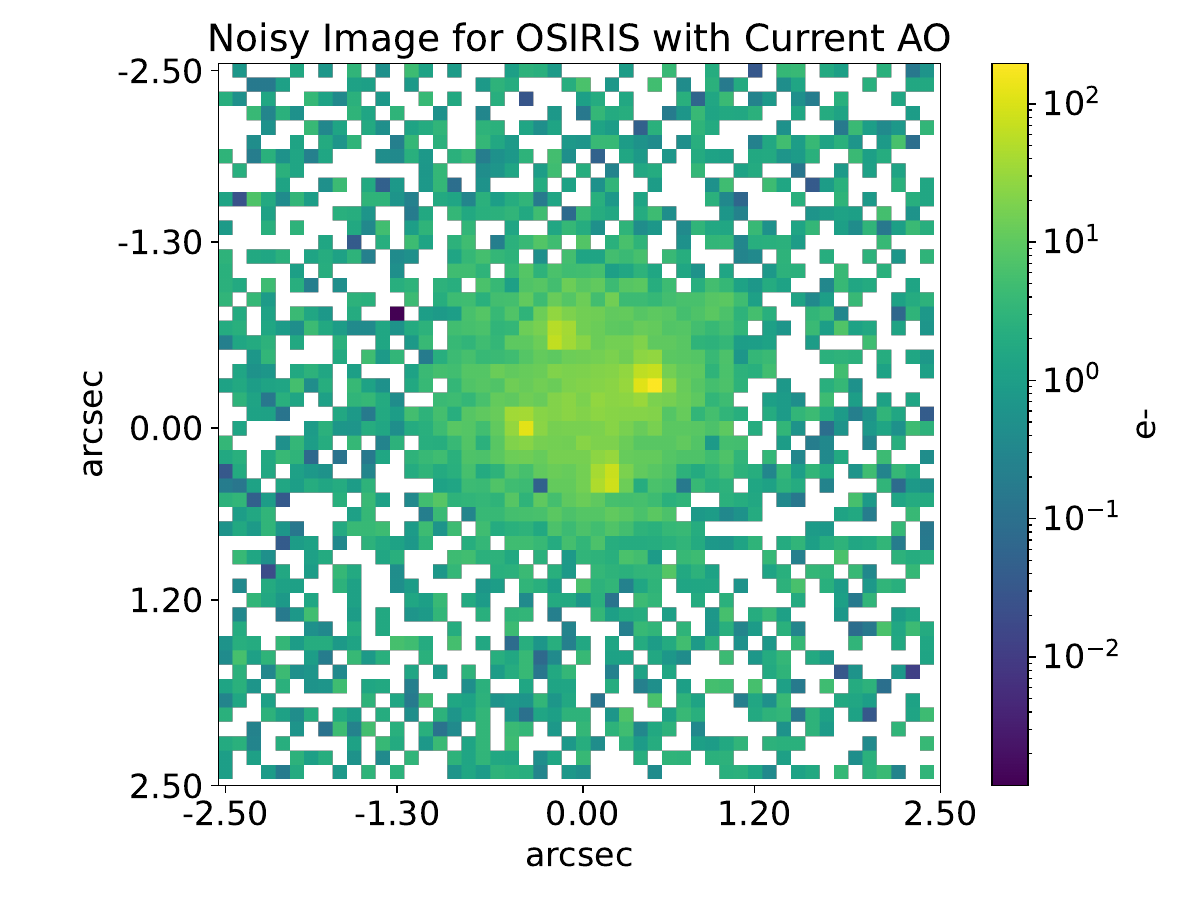} &
			\hspace{-9.2mm}
			\includegraphics[width=0.3\textwidth]{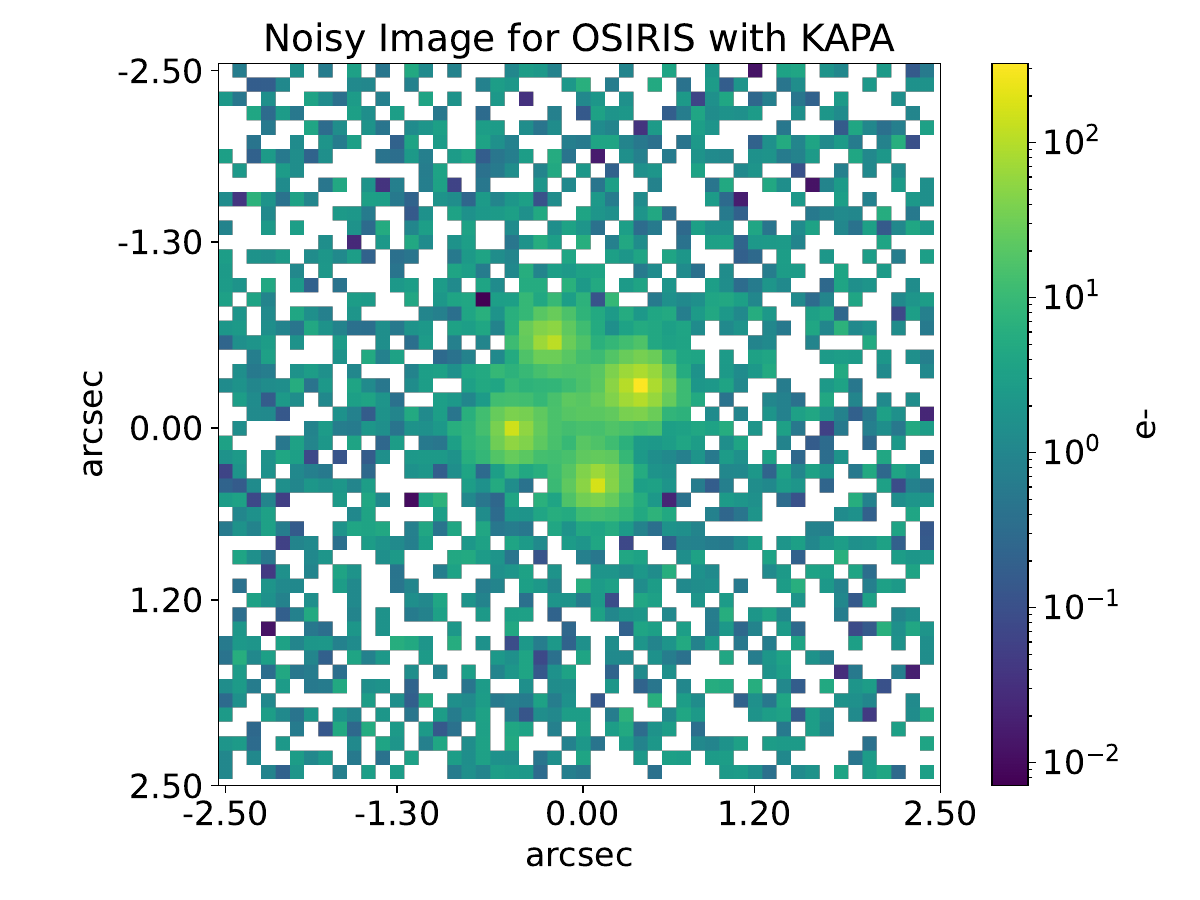} &
			\hspace{-9.2mm}
			\includegraphics[width=0.3\textwidth]{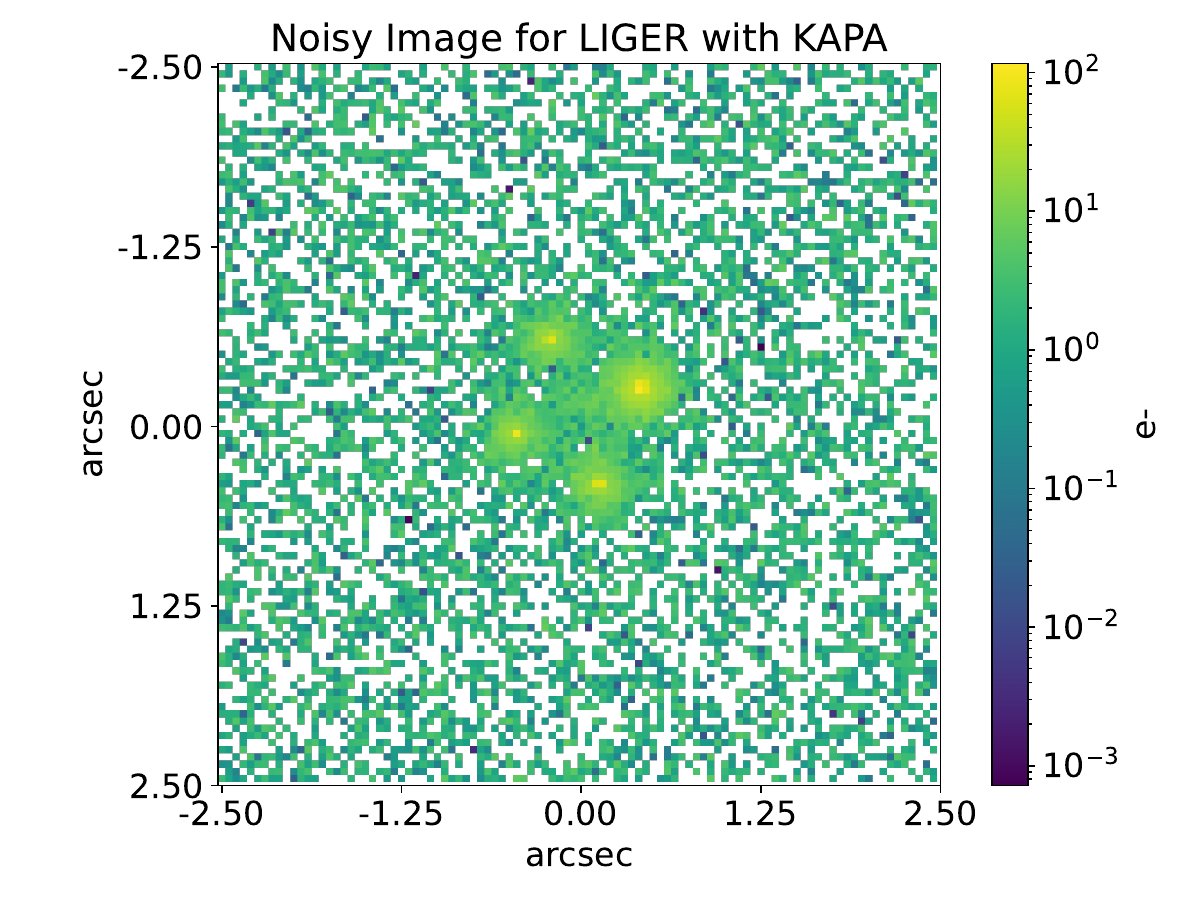} &
			\hspace{-9.2mm}
			\includegraphics[width=0.3\textwidth]{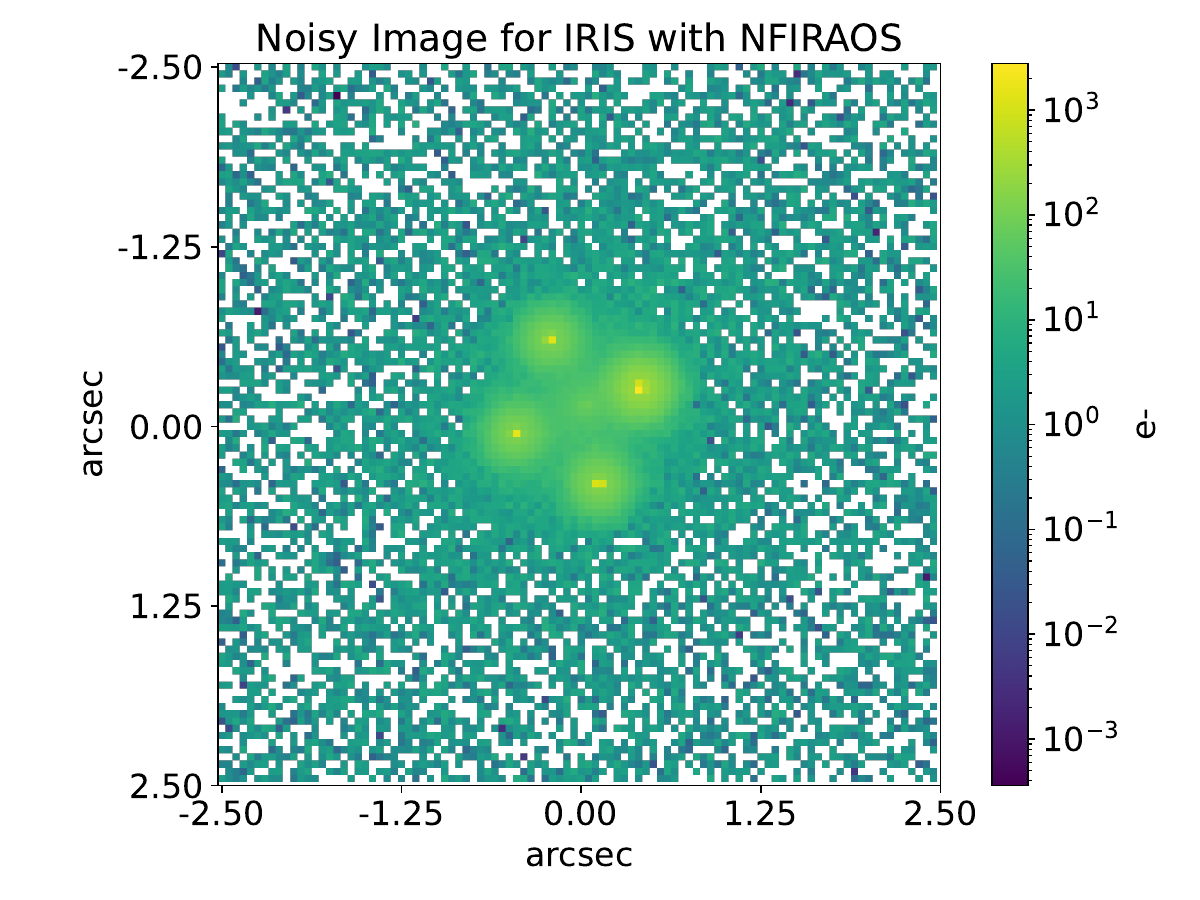} \\
			\hspace{-12mm}
			\includegraphics[width=0.3\textwidth]{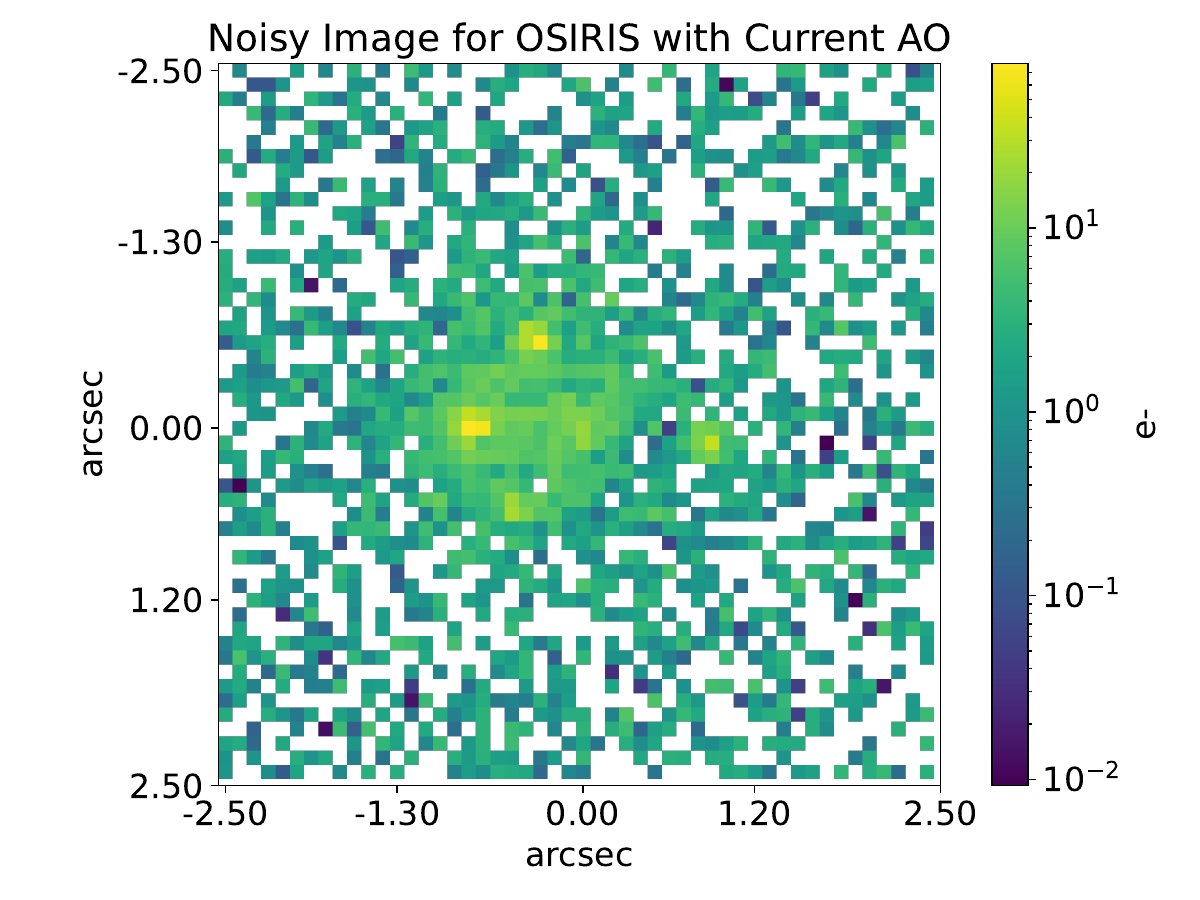} &
			\hspace{-9.2mm}
			\includegraphics[width=0.3\textwidth]{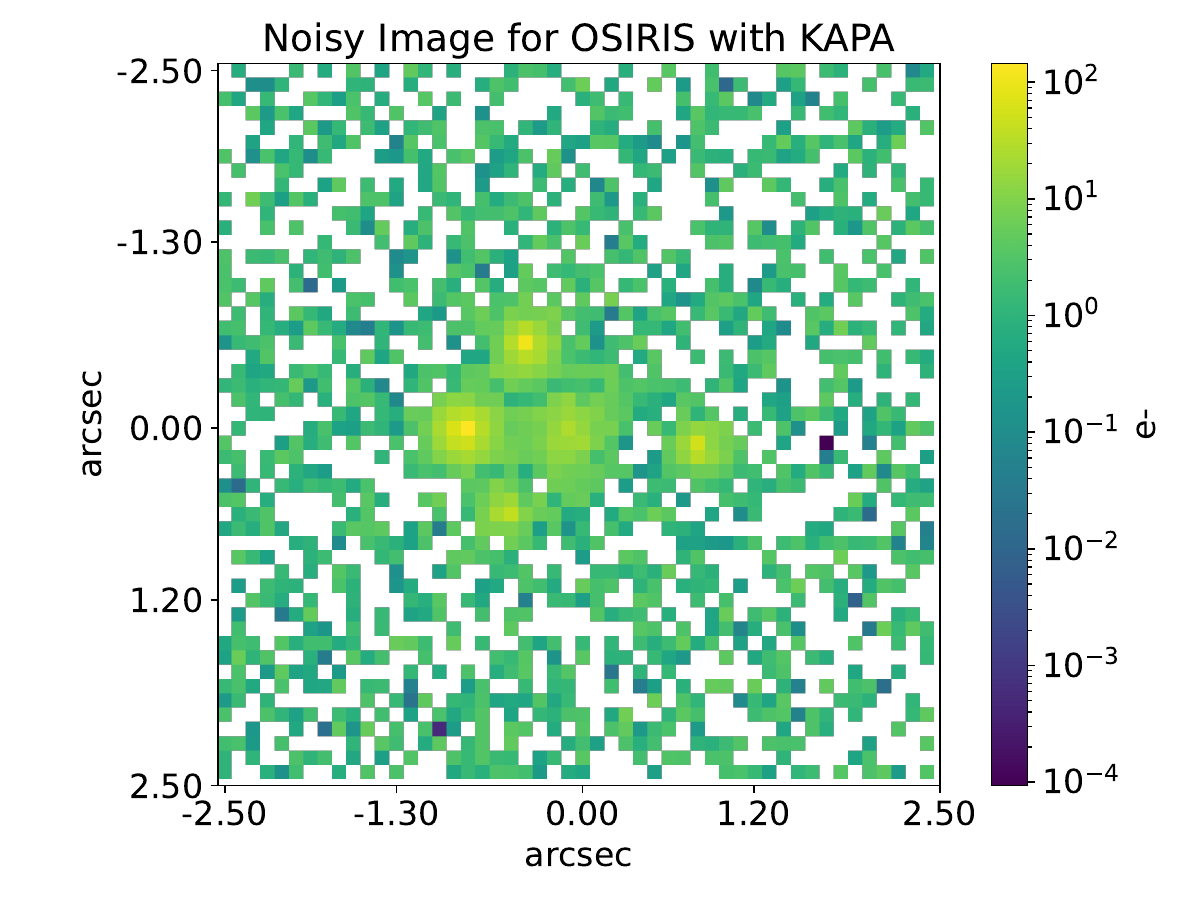} &
			\hspace{-9.2mm}
			\includegraphics[width=0.3\textwidth]{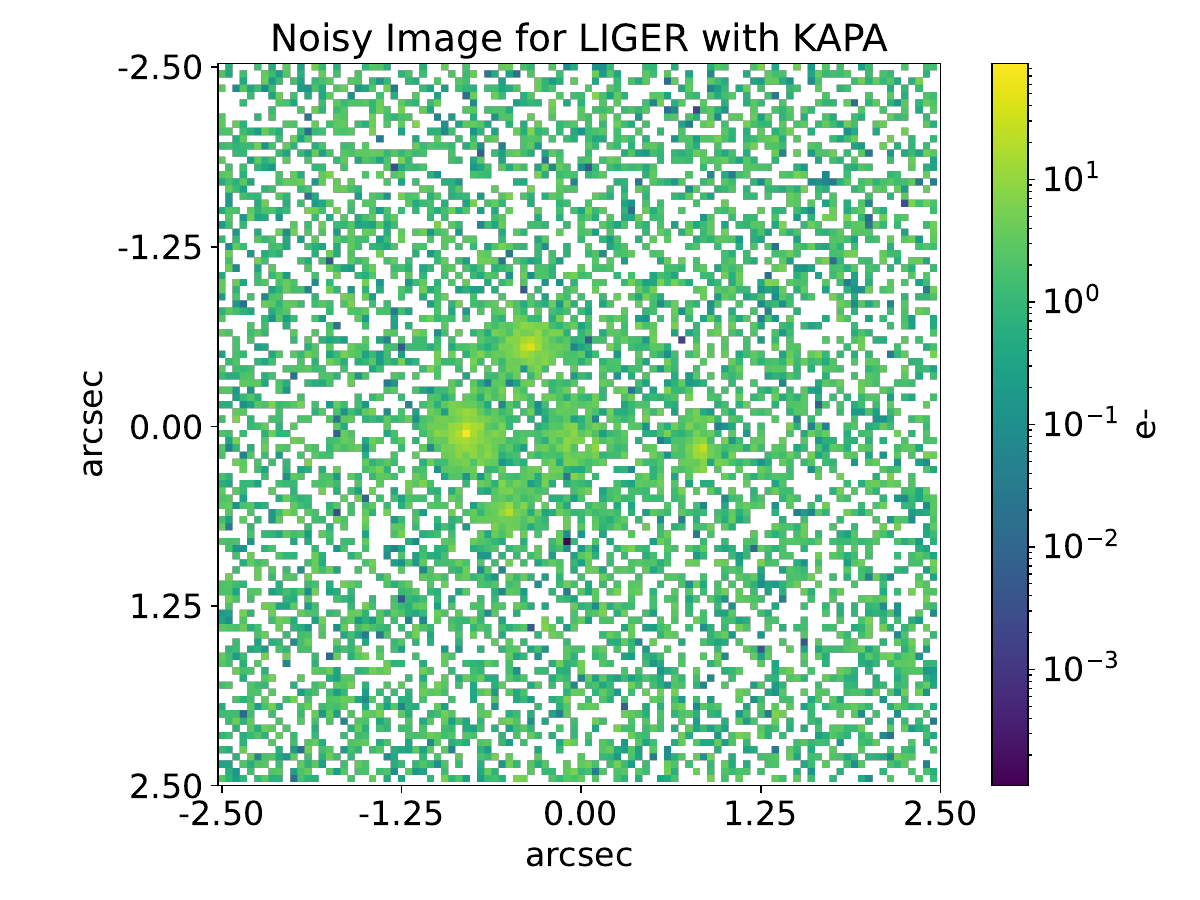} &
			\hspace{-9.2mm}
			\includegraphics[width=0.3\textwidth]{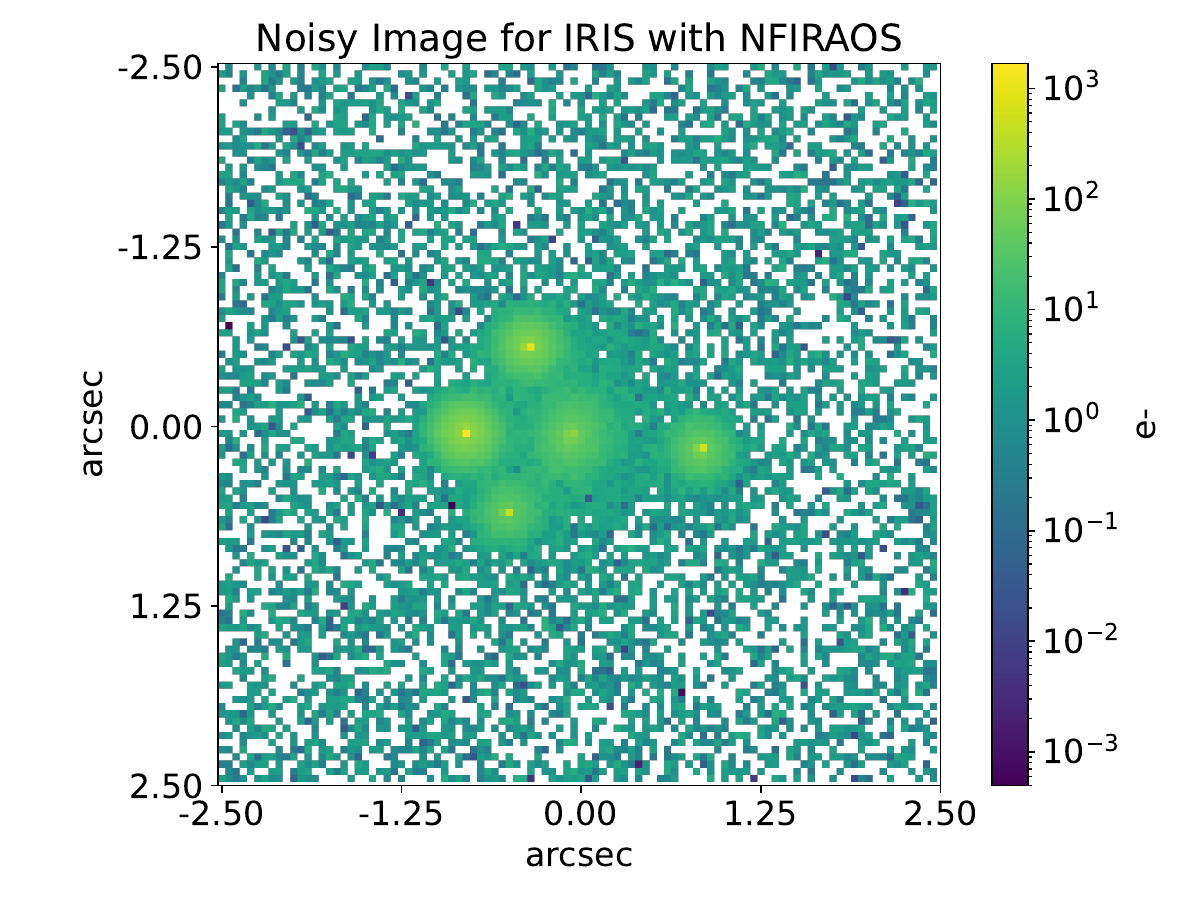} \\
			\hspace{-12mm}
			\includegraphics[width=0.3\textwidth]{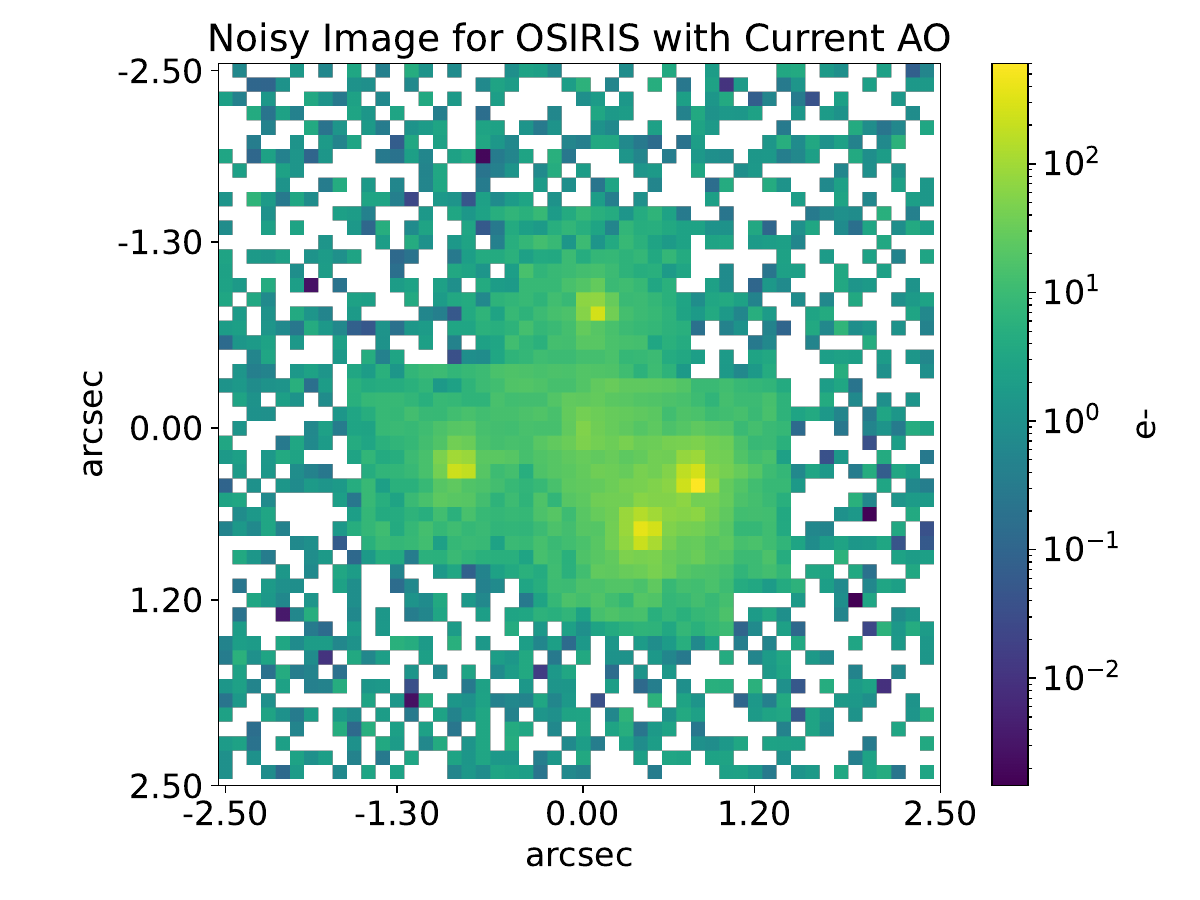} &
			\hspace{-9.2mm}
			\includegraphics[width=0.3\textwidth]{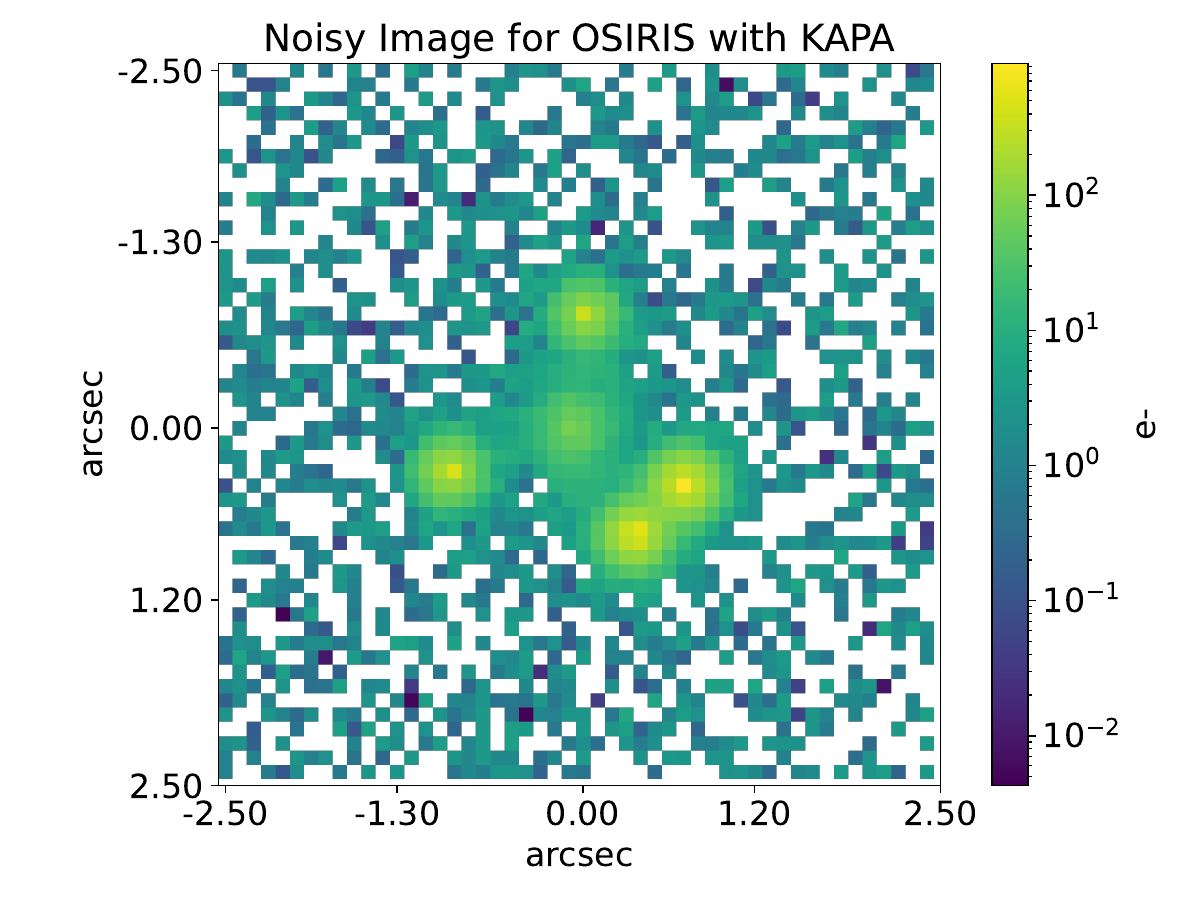} &
			\hspace{-9.2mm}
			\includegraphics[width=0.3\textwidth]{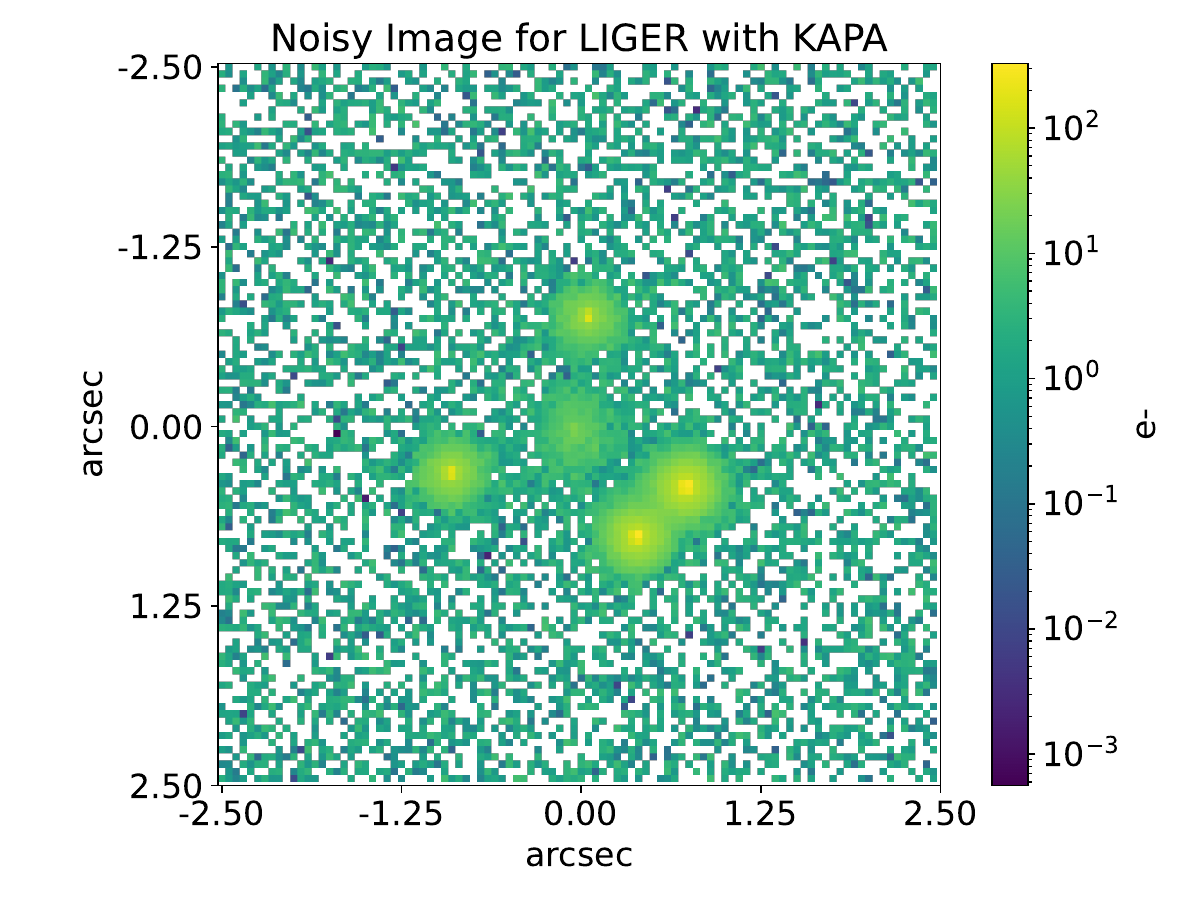} &
			\hspace{-9.2mm}
			\includegraphics[width=0.3\textwidth]{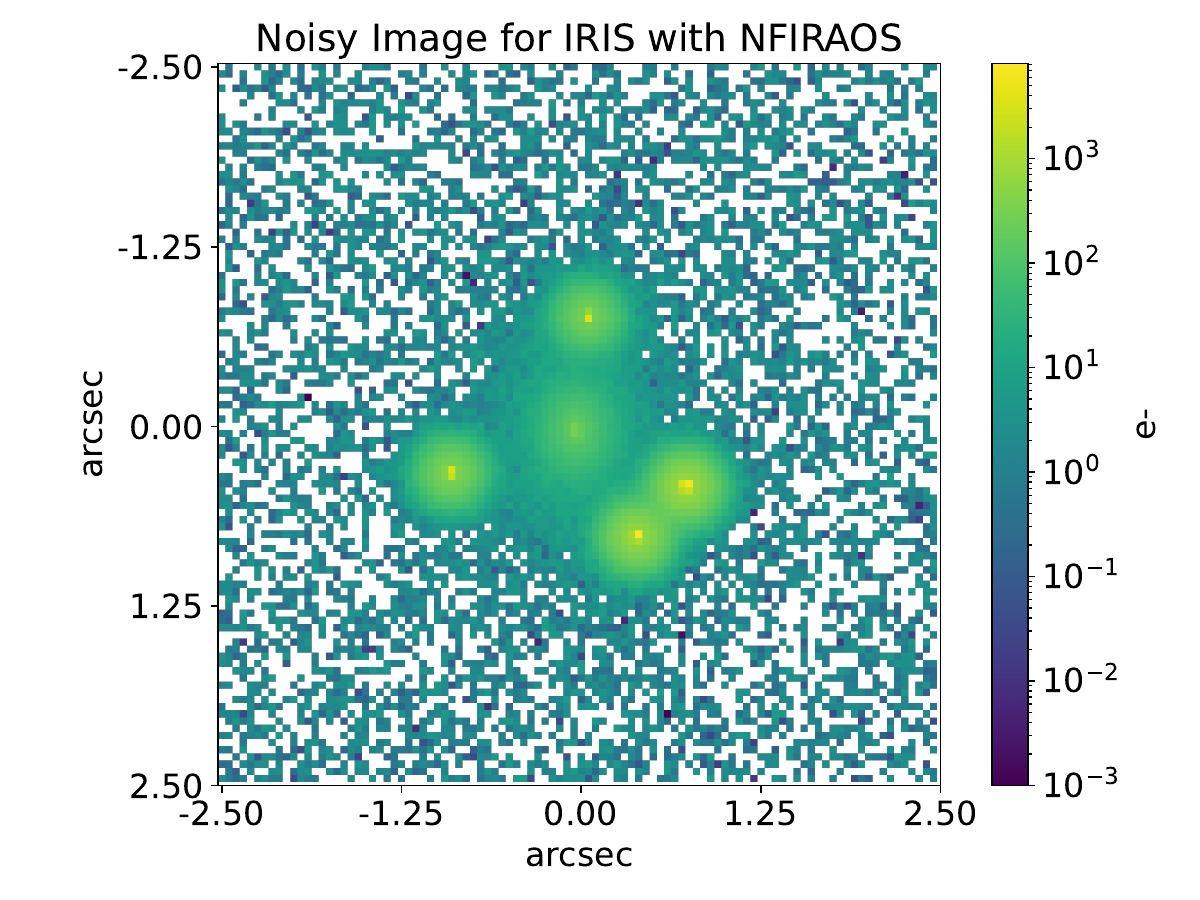} \\
		\end{tabular}
			\caption{Simulations of a cross, cusp and fold system as they would appear through the 4 different observatory configurations: OSIRIS+current adaptive optics system, OSIRIS with KAPA, LIGER with KAPA, and IRIS with NFIRAOS, with the noise levels corresponding to each system overlayed as well, presented in log scale for the effect of the PSF to be visible.}{ \label{fig:noisy_simulated_images_log}}
	\end{center}
\end{figure*}

\section{Parameter Inference}
\begin{figure*}
	\begin{center}
		\begin{tabular}{cccc}
			\hspace{-5.20mm}
			\includegraphics[width=0.5\textwidth]{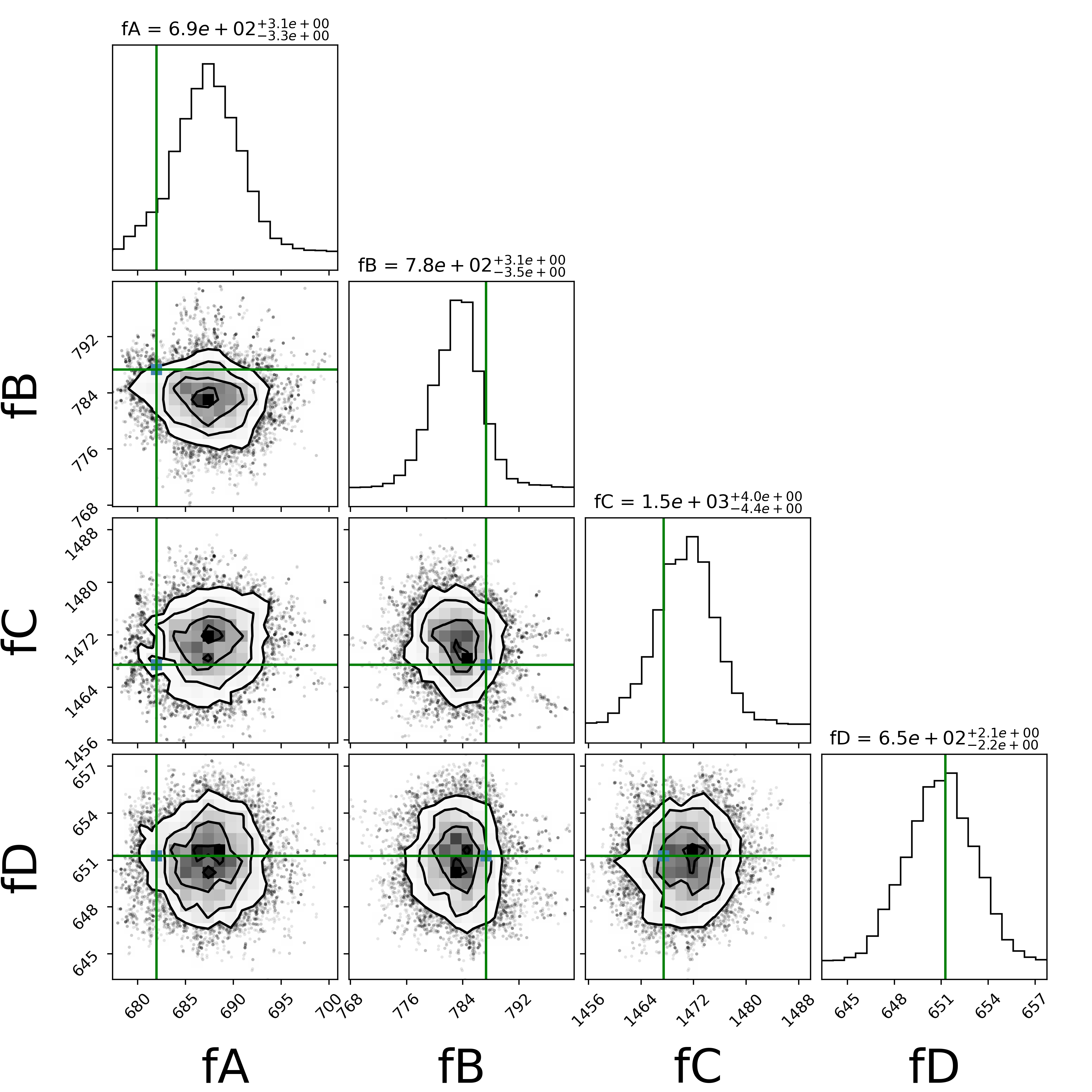}
			&
			\hspace{-4.70mm}
			\includegraphics[width=0.5\textwidth]{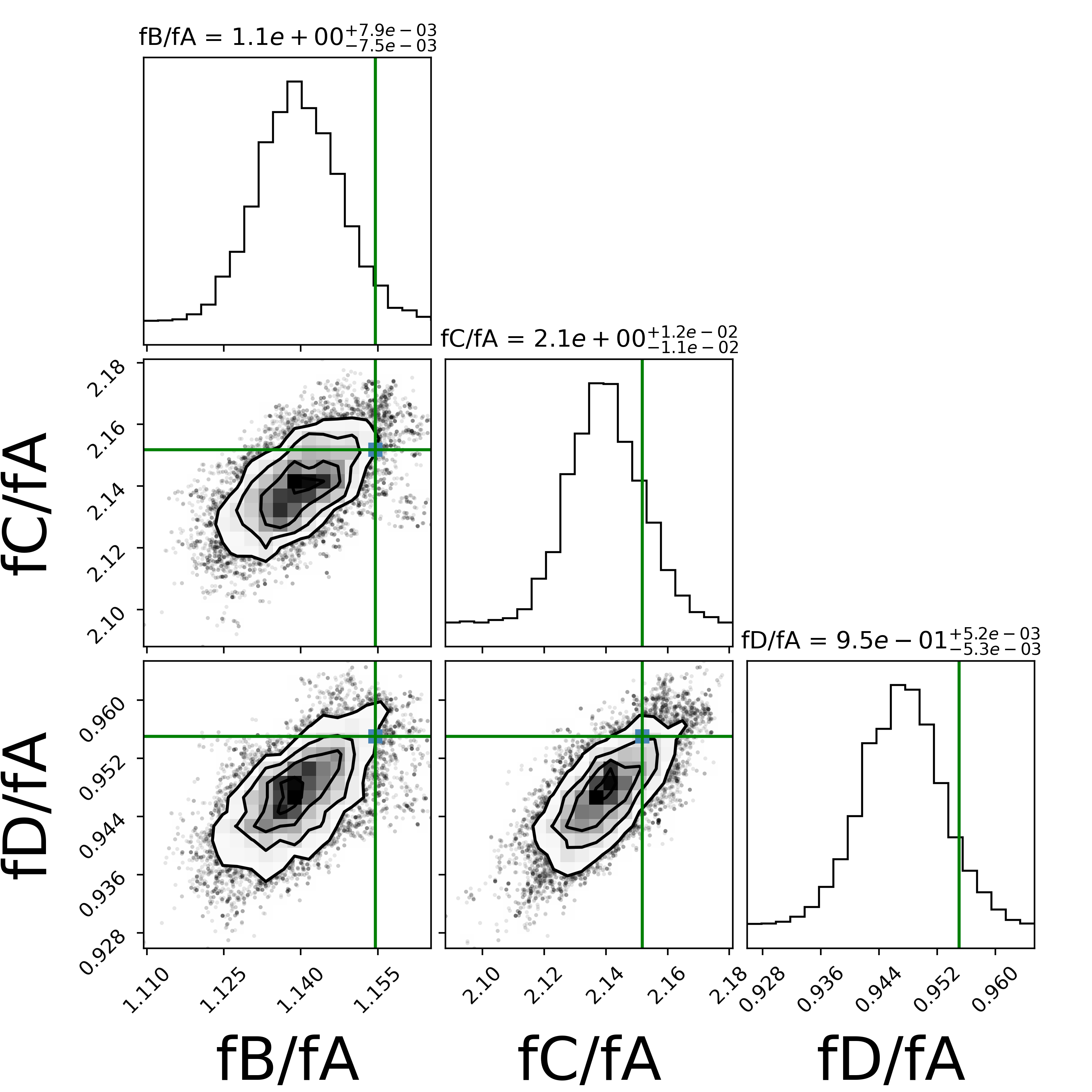}
		\end{tabular}
		\caption{The posteriors for the flux and flux ratio image measurements for the OSIRIS+KAPA configuration, with a galaxy light profile fit, for a Strehl of 0.5 and seeing FWHM of 0\farcs3. The vertical green lines indicate the true values of the fluxes and flux ratios, respectively. The posterior is Gaussian and the original flux ratios are properly recovered. See Figs. \ref{fig:master_known_psf_error_analysis}, 
  and \ref{fig:bar_plot} for more vizualizations of the errors. }
			{ \label{fig:flux_posterior}}
	\end{center}
\end{figure*}

\begin{figure*}
	\begin{center}
		\begin{tabular}{c}
			\includegraphics[width =\textwidth]{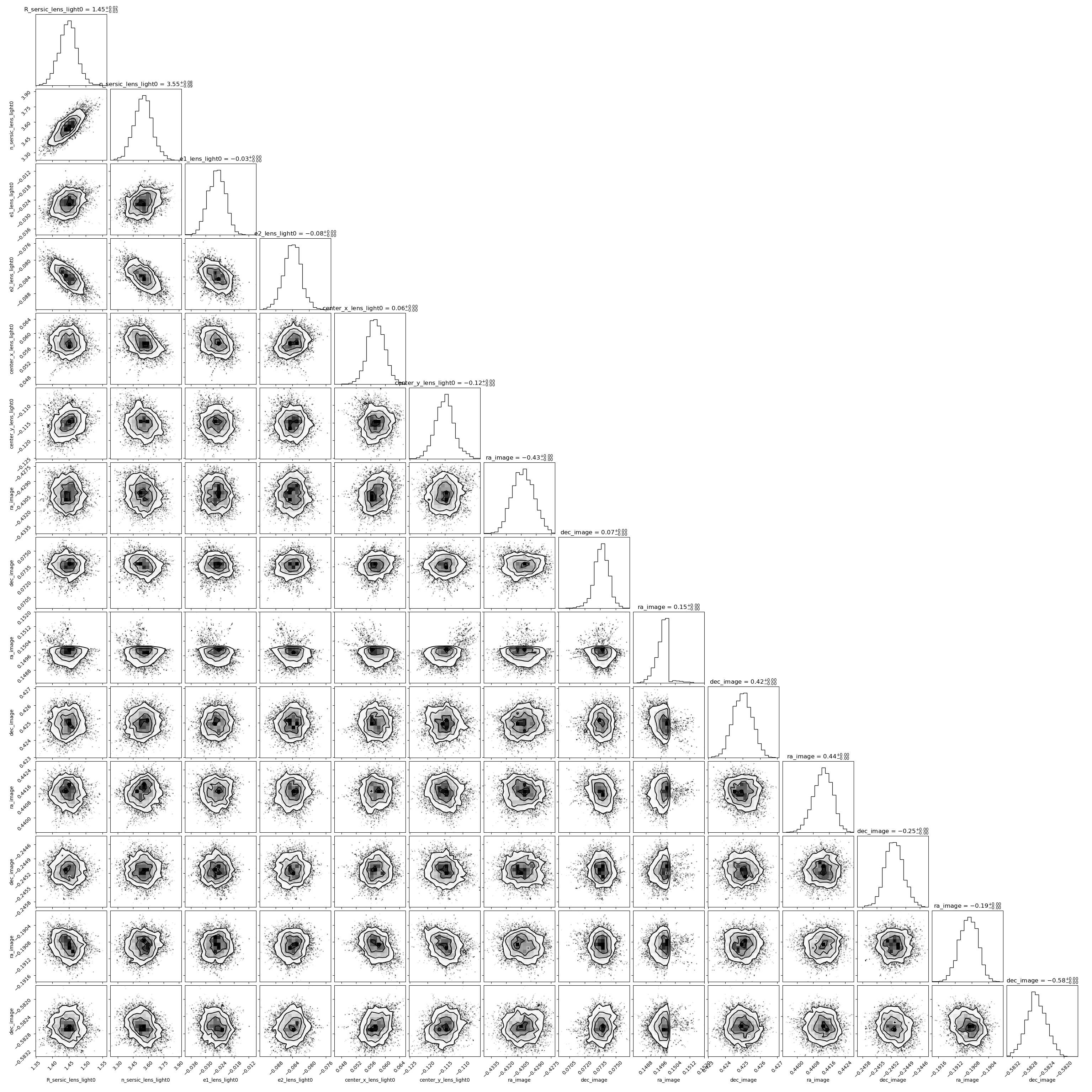}\\

		\end{tabular}
		\caption{The posteriors for MCMC parameters for the OSIRIS+KAPA configuration, with a galaxy light profile fit, for a Strehl of 0.5 and seeing FWHM of 0\farcs3. The first 6 parameters correspond to the lens light model and the last 8 to the positions of the images. }{ \label{fig:position_posterior}}
	\end{center}
\end{figure*}

In Fig \ref{fig:flux_posterior} we show an example of the fluxes and the flux ratios resulting from the MCMC posterior, for the OSIRIS and KAPA configuration, for a 600s exposure, Strehl of 0.5 and seeing FWHM of 0\farcs3. The PSF is assumed to be known, the lensed galaxy is modeled. We see that the marginalised posterior probability distributions are Gaussian, and the original fluxes and flux ratios are properly recovered. The astrometry for the image positions, together with the inferred parameters of the lensing galaxy, are shown in Fig. \ref{fig:position_posterior}. 

The posteriors for the flux and flux ratio image measurements for the OSIRIS+KAPA configuration, with a galaxy light profile fit, for a Strehl of 0.5 and seeing FWHM of 0\farcs3.




\bsp	
\label{lastpage}
\end{document}

%% file: instrument_specs.tex
\begin{table*}
\centering
\caption{Observational facilities (telescopes, detectors, and adaptive optics systems) considered in this work. The information is provided for the spectrometers (not imager) systems.}
\label{table:instrument_specs}
\begin{tblr}{
  width = \linewidth,
  row{3} = {c},
  row{4} = {c},
  cell{1}{2} = {c},
  cell{1}{3} = {c},
  cell{1}{4} = {c},
  cell{1}{5} = {c},
  cell{1}{6} = {c},
  cell{1}{7} = {c},
  cell{2}{1} = {r=3}{},
  cell{2}{2} = {r=2}{c},
  cell{2}{3} = {c},
  cell{2}{4} = {r=2}{c},
  cell{2}{5} = {r=2}{c},
  cell{2}{6} = {r=2}{c},
  cell{2}{7} = {r=2}{c},
  cell{5}{2} = {c},
  cell{5}{3} = {c},
  cell{5}{4} = {c},
  cell{5}{5} = {c},
  cell{5}{6} = {c},
  cell{5}{7} = {c},
  vline{1,8} = {1-5}{},
  vline{2} = {2-4}{dashed},
  vline{8} = {4}{},
  hline{1-2,5-6} = {-}{},
  hline{4} = {2-7}{dashed},
  hline{5} = {1-7}{dashed}
}
Telescope & Detector & $\begin{array}{c}\text{Adaptive Optics} \\ \text{System} \end{array}$ & Filter              & $\begin{array}{c}\text { Detector Dark Current } \\\left(\mathrm{e}^{-} / \text {pixel/s }\right)\end{array}$ & $\begin{array}{c}\text { Readout Noise } \\\left(\mathrm{e}^{-} / \text {pixel }\right)\end{array}$ &  $\begin{array}{c}\text { Pixel Size } \\(\operatorname{arcsec})\end{array}$ \\
KECK 10m  & OSIRIS   & Current AO & $H_{bb}$ & 0.025     & 22.   &    0.1   \\
          &          & KAPA   &  $H_{bb}$   &     &     &         \\
          & LIGER    & KAPA   &   $H_{bb}$  &  0.002 & 5.0   & 0.05   \\
TMT  30m  & IRIS     & NFIRAOS  & $H_{bb}$   & 0.002  & 5.0  &     0.05            
\end{tblr}
\end{table*}